\documentclass[conference]{IEEEtran}
\IEEEoverridecommandlockouts

\usepackage[ruled,vlined,linesnumbered]{algorithm2e}
\usepackage{cite}
\usepackage{amsmath,amssymb,amsfonts}
\usepackage{mathtools}
\usepackage{algorithmic}

\usepackage{graphicx}
\usepackage{textcomp}
\usepackage{xcolor}
\usepackage{enumitem}
\usepackage{booktabs}

\usepackage[caption=false,font=footnotesize]{subfig}

\usepackage{multirow}
\usepackage[hidelinks]{hyperref}

\def\BibTeX{{\rm B\kern-.05em{\sc i\kern-.025em b}\kern-.08em
    T\kern-.1667em\lower.7ex\hbox{E}\kern-.125emX}}

\begin{document}

\title{LA‑IMR: Latency‑Aware, Predictive In‑Memory Routing \& Proactive Autoscaling for Tail‑Latency‑Sensitive Cloud Robotics
\thanks{.}
}

\author{
    \IEEEauthorblockN{Eunil Seo\IEEEauthorrefmark{1}, Chanh Nguyen\IEEEauthorrefmark{1}, Erik Elmroth\IEEEauthorrefmark{1}}
    \IEEEauthorblockA{\IEEEauthorrefmark{1}Department of Computing Science, Ume{\aa} University, 90187 Ume{\aa}, Sweden\\
    Email: \{eunil.seo, chanh, elmroth\}@cs.umu.se}
}

\maketitle

\begin{abstract}
Hybrid cloud–edge infrastructures now support latency-critical workloads ranging from autonomous vehicles and surgical robotics to immersive AR/VR. However, they continue to experience crippling \emph{long-tail} latency spikes whenever bursty request streams exceed the capacity of heterogeneous edge and cloud tiers. To address these \emph{long-tail} latency issues, we present Latency-Aware, Predictive In-Memory Routing and Proactive Autoscaling (LA‑IMR). This control layer integrates a closed-form, utilization-driven latency model with event-driven scheduling, replica autoscaling, and edge-to-cloud offloading to mitigate 99th-percentile (P99) delays. Our analytic model decomposes end-to-end latency into processing, network, and queuing components, expressing inference latency as an affine power-law function of instance utilization. Once calibrated, it produces two complementary functions that drive: (i) millisecond-scale routing decisions for traffic offloading, and (ii) capacity planning that jointly determines replica pool sizes. LA‑IMR enacts these decisions through a quality-differentiated, multi-queue scheduler and a custom-metric Kubernetes autoscaler that scales replicas proactively—before queues build up—rather than reactively based on lagging CPU metrics. Across representative vision workloads (YOLOv5m and EfficientDet) and bursty arrival traces, LA‑IMR reduces P99 latency by up to 20.7\% compared to traditional latency-only autoscaling, laying a principled foundation for next-generation, tail-tolerant cloud–edge inference services.
\end{abstract}

\begin{IEEEkeywords}
hybrid cloud–edge computing, tail‑latency mitigation, predictive autoscaling, in‑memory routing, SLO‑aware scheduling, edge offloading, latency modeling, microservice architecture, Kubernetes HPA, 99th‑percentile (P99) latency.
\end{IEEEkeywords}

\section{Introduction}
Mitigating long-tail latency in hybrid cloud–edge systems is increasingly critical as these environments scale in complexity~\cite{suresh2015c3, dean2013tail}. Workloads span a heterogeneous continuum—from high-accuracy, resource-intensive cloud models to lightweight, low-latency edge models—making it challenging to meet strict worst-case latency targets like P99~\cite{kannan2019grandslam}. While average latency is well studied, rare yet severe spikes can deteriorate user trust and degrade performance~\cite{wang2022edge, abouaomar2022resource}. These long-tail anomalies are particularly harmful in mission-critical applications such as autonomous vehicles, industrial automation, and medical robotics~\cite{bai2019latency, zhang2024dependency}.

\begin{figure}[ht]
    \centering
    \includegraphics[width=1\columnwidth]{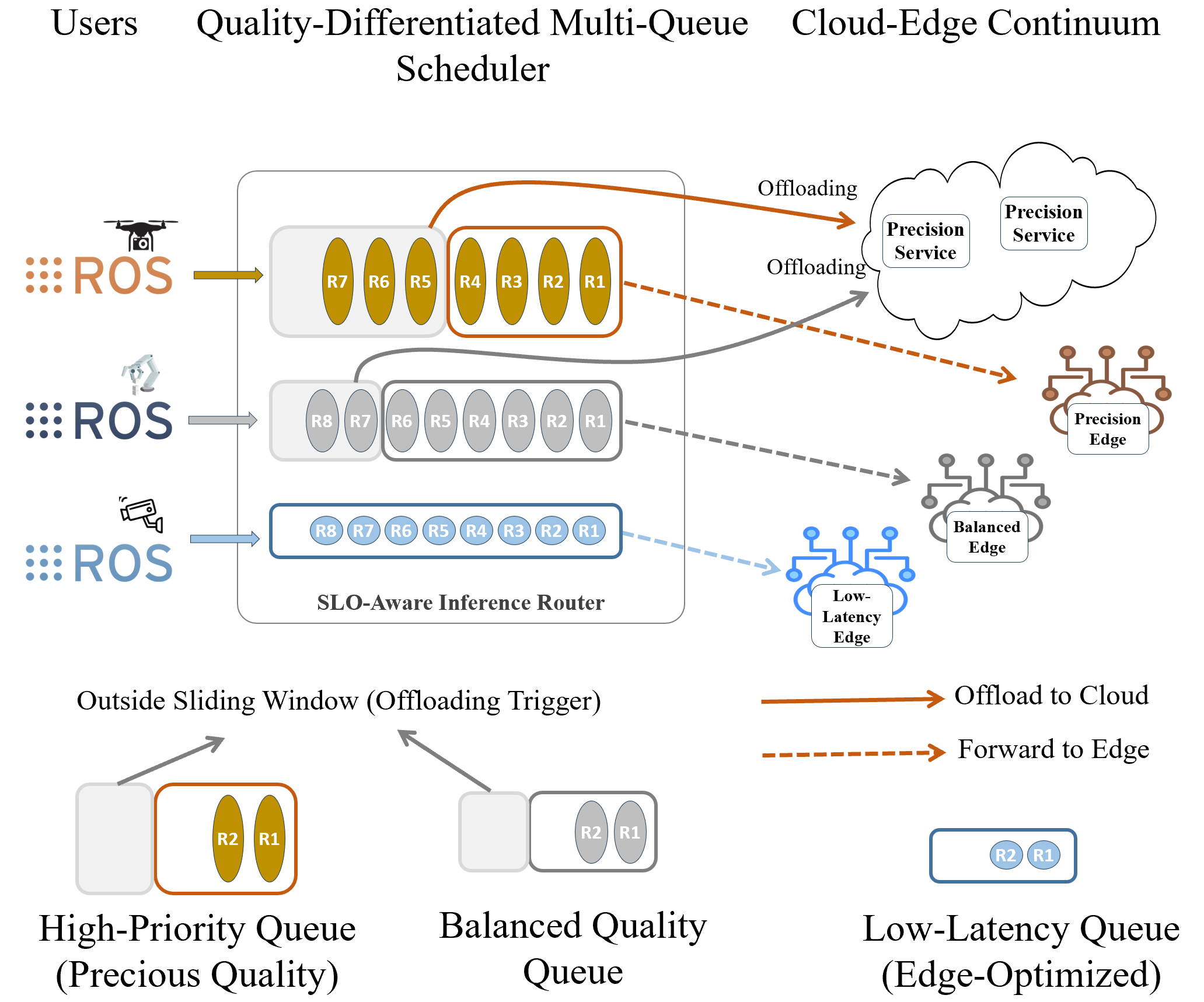}
    \caption{LA‑IMR: an in‑memory SLO‑aware controller that routes requests across edge–cloud tiers.}
    \label{fig:architecture}
\end{figure}

This work introduces an \textbf{SLO‑aware, in‑memory control loop} that sits inside a tiered microservice architecture.  As illustrated in Fig.~\ref{fig:architecture}, inference is decomposed into lightweight edge models for low-latency requests and high‑accuracy cloud models for precision tasks.  The controller steers each request to the corresponding tier that can enable its SLO—or pre‑emptively off‑loads it upstream when an early‑warning spike is detected—thereby suppressing tail‑latency anomalies.

The LA-IMR router maintains all telemetry data—including the EWMA-smoothed arrival rate, queue depth, and utilization—in process memory, updating it with every request. At the first sign of a load spike, it either (1) scales replicas using Kubernetes HPA or (2) defers excess traffic to the cloud or a faster upstream tier, thereby mitigating P99 latency. Since no external cache (e.g., Redis~\cite{redis}) is involved, these decisions incur only microseconds of access time, enabling millisecond-level responses that are essential for handling bursty, latency-critical workloads in highly latency-sensitive scenarios~\cite{zhang2021ultra, liu2018offloading}.

We introduce a closed-form, end-to-end latency model that predicts the response time of any inference request routed through an edge–cloud continuum. The model decomposes latency into three components: (i) an inference-processing term that scales with instance utilization according to an affine power law, (ii) a task-agnostic network round-trip time (RTT), and (iii) an analytically derived M/M/c queuing delay. Calibrated using only three parameters per hardware tier—the model’s reference latency $L_m$, the hardware speed-up $S_{m,i}$, and a super-linearity exponent $\gamma$—the model effectively captures how latency increases under bursty loads, heterogeneous hardware, and varying replica counts. Extensive measurements demonstrate that this single equation tracks observed latencies within a few percent across a wide operational range, enabling the runtime to anticipate SLO violations and to support proactive routing, autoscaling, and offloading decisions throughout the system.

By combining proactive latency‑spike detection with utilization‑driven autoscaling and real‑time, in‑memory telemetry, LA‑IMR adapts within milliseconds to traffic bursts or faults, shrinking long‑tail latency and bolstering overall system reliability. Building on this foundation, our work contributes:
\begin{enumerate}[leftmargin=*]
    \item \textbf{Closed‑form, dual‑purpose latency model}. We derive a single analytic equation that decomposes end‑to‑end delay into processing, network, and \(\mathrm{M/M/c}\) queuing terms and captures super‑linear contention with one exponent. Two complementary instantiations—fixed‑replica \(g_{m,i}(\boldsymbol\lambda)\) and fixed‑traffic \(g_{m,i}(N_{m,i})\)—drive millisecond‑scale routing decisions as well as slower capacity‑planning optimisation.
    \item \textbf{Tail‑aware, quality‑stratified request routing}.  LA‑IMR is an event‑driven, in‑memory controller that predicts imminent P99 breaches, routes requests across latency/accuracy‑differentiated queues, and pre‑emptively offloads traffic or instantiates replicas before long‑tail spikes materialise.          
    \item \textbf{Proactive autoscaling from model‑predicted metrics}. By exporting the required replica count, as computed from the latency model, as a custom Kubernetes metric, the system eliminates the 60–120s lag and oscillations of CPU‑driven HPA, scaling just‑in‑time to keep queues short.
    \item \textbf{Empirical gains on bursty vision workloads}. Experiments with YOLOv5m and EfficientDet under bursty traces show that LA‑IMR trims P99 latency by up to 20.7\%, confirming the practical impact of the theory‑driven design.
\end{enumerate}

This work presents LA‑IMR, an SLO‑aware control layer that proactively routes, scales, and off‑loads AI inference to suppress long‑tail latency in hybrid edge–cloud systems: Section~\ref{sec:2} surveys related tail‑latency and resource‑allocation work; Section~\ref{sec:3} formalises a closed‑form latency model and optimisation framework; Section~\ref{sec:4} details the LA‑IMR architecture—multi‑queue scheduler, predictive router, and custom‑metric Kubernetes autoscaler; Section~\ref{sec:5} demonstrates up to 20.7\% P99‑latency reductions under bursty loads; and Section~\ref{sec:conclusion} sketches future paths toward resilient, tail‑tolerant AI services.

\section{Related work} \label{sec:2}

\subsection{Tail Latency Mitigation}
Early work formulated long‑tail delay as a fault to be masked rather than eliminated. \textit{C3} steers reads toward the fastest‑responding replica while throttling overly aggressive clients, trimming the 99th‑percentile (P99) latency without global coordination~\cite{suresh2015c3}. In large fan‑out services, Dean and Barroso advocate hedged or tied requests so that a straggler no longer determines end‑to‑end response time~\cite{dean2013tail}. These “speculative” techniques are powerful, yet they operate \emph{after} latency inflation has already begun and provide little guidance on resource sizing.

More recent systems push decision making inside the service graph. \textit{GrandSLAm} predicts per‑stage completion times, then reorders or batches microservice calls to respect job‑level SLAs while sustaining throughput~\cite{kannan2019grandslam}. Complementary edge–cloud frameworks model DNN inference pipelines and split them across heterogeneous hardware to keep worst‑case delay bounded under fluctuating bandwidth and load~\cite{wang2022edge, liang2023model, abouaomar2022resource, rao2021eco}. These platforms, however, typically rely on threshold‑based autoscalers or coarse queue metrics, which react only after the utilisation spike is visible.

Offloading augments local scheduling by opportunistically handing work to faster or less‑loaded tiers. Queue‑length‑aware fog dispatchers~\cite{hwang2021queue}, partial offload optimisers that jointly minimise energy and latency~\cite{ahmad2022partial}, and reinforcement‑learning controllers for mobile MEC~\cite{zhang2021ultra} all illustrate the benefit of decoupling execution sites. Extensions with IRS‑assisted channels~\cite{bai2019latency}, dependency‑aware scheduling~\cite{zhang2024dependency}, or sub‑task partitioning~\cite{liu2018offloading} further reduce tail variance, yet they rarely couple the offloading trigger to a predictive latency model that spans processing, network, and queueing effects.

LA‑IMR bridges these gaps by (i) deriving a closed‑form, utilisation‑driven latency law that can be evaluated in microseconds, (ii) embedding that model in an event‑driven multi‑queue router that makes per‑request routing and offloading decisions, and (iii) exporting a custom metric to Kubernetes so replicas scale \emph{proactively}, before queues build.

\subsection{Autoscaling and Resource Management}

Early autoscalers integrated queueing theory with threshold- or model-driven control. Gandhi et al.'s adaptive model-driven autoscaler reduces SLA violations by accurately forecasting required capacity~\cite{gandhi2014autoscaling}. Cluster managers implemented isolation mechanisms similar to Borg's priority and quota controls for latency-critical jobs~\cite{verma2015large}. With microservices, the focus has shifted to replicas for each service and resource management that is aware of Service Level Objectives (SLOs). Kubernetes's Horizontal Pod Autoscaler (HPA) scales pods based on resource or custom metrics~\cite{kubernetes_hpa_2025}, while FIRM and Sinan utilize telemetry-driven or learned models for controlling microservice resources~\cite{qiu2020firm, zhang2021sinan}. LA-IMR utilizes a closed-form, utilization-driven model to forecast SLO pressure, trigger event-driven routing, and provide custom replica targets to Kubernetes before tail latency becomes an issue.


\subsection{In-Memory Processing and Dynamic Routing}

Prior efforts to address tail-latency spikes have focused on three key factors---in-memory state, dynamic offloading, and fine-grained replica control---yet none has unified them under a single predictive model, as LA-IMR does. Heracles partitions cores, memory, and caches to confine data-center interference, but it cannot forecast request-level SLO breaches in advance~\cite{Lo2015Heracles}. FaRM and FASTER lower the latency floor through RDMA-backed and hybrid in-memory key--value stores, respectively, but both leave queue buildup to the application layer rather than modeling it as a first-class control signal~\cite{Dragojevic2014FaRM, Chandramouli2018FASTER}. A closer line of work couples dynamic placement with edge--cloud inference. Neurosurgeon partitions DNN execution between mobile devices and the cloud to reduce response time~\cite{Kang2017Neurosurgeon}, while Jellyfish considers inference serving with end-to-end latency SLOs over dynamic edge networks~\cite{9984750}.

LA‑IMR extends this lineage by (i) holding all routing telemetry in process memory for sub‑millisecond decisions, (ii) predicting per‑replica latency with a calibrated utilization model so replicas can be prepared to begin \textit{before} queues surge, and (iii) integrating offloading, routing, and autoscaling into one event‑driven loop.

\subsection{Cloud Robotics}
Object detection in cloud robotics balances accuracy against latency amid bursty, location‑dependent demand. High‑precision detectors such as Faster and Mask R‑CNN excel in the cloud but suffer multi‑hundred‑millisecond delays when GPU queues saturate~\cite{ren2015faster}. Conversely, edge‑optimised one‑stage models (e.g., EfficientDet, MobileNet‑SSD) return sub‑50ms results on lightweight devices, yet their lower mAP limits use in safety‑critical contexts~\cite{farhadi2018yolov3}. Foundational work—from R‑CNN through PASCAL VOC—still guides model‑selection policy~\cite{szegedy2013deep,everingham2010pascal}, and Huang~\textit{et al.} showed that no single network meets all SLOs across traffic regimes~\cite{huang2017speed}.

Traditional cloud‑edge schedulers rely on coarse utilisation thresholds, scaling only after queues build, which drives P99 latencies to exceed the mean by more than 5× during bursts~\cite{suresh2015c3,dean2013tail}. LA‑IMR embeds a closed‑form, utilisation‑aware latency model and a quality‑differentiated multi‑queue scheduler directly in the inference path. It scales replicas or off‑loads traffic \emph{proactively}, maintaining task‑level P99 within SLOs despite workload spikes, fluctuating RTT, and hardware heterogeneity—offering a principled antidote to tail‑latency anomalies in cloud‑robotic perception.

\section{System Model and Problem Formulation}\label{sec:3}
Inspired by foundational research on latency modeling and resource allocation in edge–cloud environments, particularly the surveys conducted by Mao et al.\cite{mao2017survey}—we develop a closed-form latency expression specifically adapted to our multi-replica hybrid infrastructure. This formulation incorporates conventional latency factors such as computation, communication, and queuing delays, while also integrating empirical latency patterns drawn from recent experimental analyses~\cite{9984750, 10.1145/3140659.3080246}.

\subsection{\textbf{Latency Components}}
\label{subsec:latency_components}

The end-to-end latency experienced by a task
\(t\in\mathcal{T}\) is the sum of processing, network, and
queuing delays:
\begin{equation}
\label{eq:latency_components}
  L_t \;=\; L_{m,i}^{\text{infer}}
        \;+\; D_{t, i}^{\text{net}}
        \;+\; Q_{t,i} ,
\end{equation}
where
\begin{itemize}
  \item \(L_{m,i}^{\text{infer}}\) — \emph{inference processing delay}: latency of
        model~\(m\) when served by instance~\(i\) under its current load;
  \item \(D_{t, i}^{\text{net}}\) — \emph{network (RTT) delay}: round-trip
        data-transfer latency between the data source and instance~\(i\);
  \item \(Q_{t, i}\) — \emph{queuing delay}: waiting time in the input queue of instance~\(i\) before execution begins.
\end{itemize}

\begin{table}[h]
\caption{Notation Table}
\centering
\begin{tabular}{ll}
\toprule
\textbf{Symbol} & \textbf{Description} \\
\midrule
$L_m$           & Mean latency of model $m$ on the CPU baseline \\
$S_{m,i}$       & Speed-up of instance $i$ for model $m$ \\
$\lambda_m$     & Aggregate arrival rate for model $m$ \\
$N_{m,i}$       & Replica count of model $m$ on instance $i$ \\
$R_m$           & Mean compute time per request of model $m$ \\
$R_i^{\max}$    & Sustainable compute budget of instance $i$ \\
$B_i$           & Background (co-tenant) load on instance $i$ \\
$\gamma>1$      & Empirical super-linear exponent \\
\bottomrule
\end{tabular}
\label{tab:notation}
\end{table}

\subsection{\textbf{Entities and Notation}}
\label{sec:latency-model}

\subsubsection{Inference tasks.}
The set of independent inference tasks is
\(\mathcal{T}=\{1,\dots,T\}\).
Each task~\(t\) specifies an accuracy requirement
\(\alpha^{\text{req}}_t\) and, optionally, a latency
service-level objective (SLO)~\(\tau_t\).

\subsubsection{Inference models.}\label{sec:inference-model}
The catalogue of ML models is
\(\mathcal{M}=\{1,\dots,M\}\).
Every model \(m\in\mathcal{M}\) is described by
\begin{itemize}
  \item \(L_m^{\text{infer}}\), steady‑state inference latency on a \emph{reference} device;
  \item $a_m\!\in[0,1]$, steady‑state accuracy;
  \item $R_m$, per‑inference resource demand (e.g., CPU‑seconds).
\end{itemize}

Instance~\(i\) offers a per-inference CPU budget
\(R_i^{\max}\). When model~\(m\) executes on instance~\(i\) its measured latency is \(L_{m,i}^{\text{infer}}\) as in Eq.~\eqref{eq:latency-util}.
We track two canonical computer-vision backbones:
\begin{align*}
  m_1 &= \text{EfficientDe},\\
  m_2 &= \text{YOLOv5m}.\\
\end{align*}
Their steady‑state inference latencies and CPU demands on a Raspberry~Pi~4 are summarised in Table~\ref{tab:model_profile}.  Note that the lightweight EfficientDe is nearly two orders of magnitude cheaper in \(R_m\) than the heavier YOLOv5m.

\begin{table}[h!]
\renewcommand{\arraystretch}{1.3}
\centering
\caption{Model profile on a reference edge instance provisioned with a Raspberry Pi 4 VM configured with 3 CPU cores. \(R_m\) is expressed in CPU-seconds per inference; \(L_m^{\text{infer}}\) is the steady-state latency (\(\pm\)~std.~err.).}
\begin{tabular}{|l|c|c|}
\hline
Model & \(L_m^{\text{infer}}\) [s] & \(R_m\) [CPU-s] \\
\hline
EfficientDet (\(m_1\)) & \(0.09\pm1.2\times10^{-3}\) & 0.10 \\
YOLOv5m (\(m_2\))       & \(0.73\pm3.0\times10^{-3}\) & 1.00 \\
\hline
\end{tabular}
\label{tab:model_profile}
\end{table}

\subsubsection{Hybrid Infrastructure as VM Instances}
We consider an edge–cloud continuum provisioned as virtual machine instances. Hence
\begin{align*}
  \mathcal{C} &= \{\,c_1,\ldots,c_{|\mathcal{C}|}\}: \text{set of \emph{cloud} VM instances},\\
  \mathcal{E} &= \{\,e_1,\ldots,e_{|\mathcal{E}|}\}: \text{set of \emph{edge} VM instances}.
\end{align*}
Their union yields the global instance set $\mathcal{I}=\mathcal{C}\cup\mathcal{E}$. Each \textbf{instance}\footnote{While instance and replica are not strictly synonymous in Kubernetes—replica denotes the desired number of concurrently running pods—this work uses the terms interchangeably for simplicity and consistency in discussion.}~$i\in\mathcal{I}$ exposes a finite resource budget $R_i^{\max}$ (e.g., CPU‑seconds, GPU‑seconds) and may be subject to an exogenous background load~$B_i$.

\subsubsection{Decision Variable}
The binary variable
\[
  x_{t,m,i} \;=\;
    \begin{cases}
      1, & \text{if task } t \text{ is executed by model } m \text{ on instance } i,\\[2pt]
      0, & \text{otherwise},
    \end{cases}
\]
encodes the joint \emph{model‑selection} and \emph{task‑placement} decision.

\subsubsection{Resource and Assignment Constraints}
Each task must be assigned exactly once:
\begin{equation}
  \sum_{m\in\mathcal{M}}\sum_{i\in\mathcal{I}} x_{t,m,i} = 1, \qquad \forall t\in\mathcal{T}.
\label{eq:assignment}
\end{equation}

The aggregate resource demand on each node shall not exceed its capacity:
\begin{equation}
  \sum_{t\in\mathcal{T}}\sum_{m\in\mathcal{M}} x_{t,m,i}\,R_m \le R_i^{\max}, \qquad \forall i\in\mathcal{I}.
\label{eq:capacity}
\end{equation}

\subsection{\textbf{Inference Processing Delay}}
\label{subsec:processing_delay}

\paragraph{\textbf{Latency as a Function of Model Size}}
Nigade~\textit{et~al.}\cite{9984750} empirically observe that the inference latency of a deep-learning model grows sub-linearly with the input batch size.  For a model $m_j$ with parameter file size $s_j$ and batch size $b$, the mean per-inference latency is
\begin{equation}
l_j(b)\;=\;\alpha\,s_j\,b^{\gamma}, \qquad 0<\gamma<1,
\label{eq:single_latency}
\end{equation}
where $\alpha$ is a hardware- and framework-dependent constant and $0<\gamma<1$ captures the sub-linear batching benefit.

\paragraph{\textbf{Utilisation-based Latency Model}}
Extending the principles of utilization-based performance modeling proposed by Wang et al.~\cite{wang2019edgeai,wang2021survey} and inspired by hardware-performance scaling insights from Jouppi et al~\cite{10.1145/3140659.3080246}, we formulate the inference latency of model~$m$ running on instance~$i$ as a function of \emph{instance utilization}~$U_i$:
\begin{equation}
L_{m,i}^{\text{infer}}(\lambda_m,N_{m,i}) \;=\; \frac{L_m}{S_{m,i}}\,
\Bigl[1 + U_i^{\gamma}\Bigr],
\label{eq:latency-util}
\end{equation}
where
\begin{itemize}
  \item $L_m$ is the single-inference latency of model~$m$ on the \emph{reference} hardware;
  \item $S_{m,i}$ is the hardware speed-up factor of instance~$i$ for model~$m$ (Table~\ref{tab:hardware_scaling});
  \item $\gamma\!\ge\!0$ controls how sharply latency rises as utilisation increases;
  \item $U_i$ is the instantaneous utilisation of instance~$i$:
        \begin{equation}
        U_i \;=\; \frac{\displaystyle \sum_{m'\in\mathcal{M}}\lambda_{m'}R_{m'} \;+\; B_i}
                         {R_i^{\max}}.
        \label{eq:utilisation}
        \end{equation}
\end{itemize}
Here:
\begin{itemize}
    \item $\lambda_{m'}$ is the arrival rate of model~$m'$.
    \item $R_{m'}$ is the resource consumption per inference for $m'$.
    \item $B_i$ denotes the background (co-tenant) load on instance~$i$.
    \item $R_i^{\max}$ is the total capacity of instance $i$.
\end{itemize}

The hardware scaling factor $S_{m,i}$ represents the hardware-dependent acceleration and is typically determined empirically. For example, Oh et al.~\cite{OH20041311} show that CPUs can be up to 20 times faster in certain scenarios, while Jouppi et al.~\cite{10.1145/3140659.3080246} report that TPUs are approximately 15 to 30 times faster than contemporary GPUs (e.g., the NVIDIA K80). However, performance may vary significantly depending on the underlying technology and specific commercial hardware.

To provide a conceptual understanding of the hardware scaling factor, this work approximates typical values, as summarized in Table~\ref{tab:hardware_scaling}.

\begin{table}[h]
\centering
\caption{Typical scalability of hardware in single- and multi-instance scenarios.}
\begin{tabular}{|l|c|}
\hline
\textbf{Hardware Type} & \shortstack{\textbf{Typical} \boldmath$S_{m,i}$} \\
\hline
CPU & 1 \\
GPU & 2–20 \\
TPU & 30–100+ \\
\hline
\end{tabular}
\label{tab:hardware_scaling}
\end{table}

\paragraph{\textbf{Affine Power-Law Form}}
During calibration we vary only the traffic of the model under study and keep co-tenancy fixed.  Writing the \emph{per-replica} arrival rate as $\tilde\lambda_{m,i}\!=\!\lambda_m/N_{m,i}$ and expanding $U_i^{\gamma}$ yields
\begin{align}
L_{m,i}^{\text{infer}}
&=\frac{L_m}{S_{m,i}}\Bigl[1 + (B_i/R_i^{\max})^{\gamma}\Bigr]   \notag\\
&\quad+ \frac{L_m}{S_{m,i}}\bigl(R_m/R_i^{\max}\bigr)^{\gamma}\,
      \tilde\lambda_{m,i}^{\gamma}                                                     \\
&\equiv \underbrace{\alpha_i}_{\text{baseline}} + 
          \underbrace{\beta_{m,i}}_{\text{slope}}\,
          \tilde\lambda_{m,i}^{\gamma},
\label{eq:inference-delay-final}
\end{align}
with
\begin{equation}
\begin{aligned}
\alpha_i \;=\; \frac{L_m}{S_{m,i}}\!\Bigl[1 + (B_i/R_i^{\max})^{\gamma}\Bigr], \\
\beta_{m,i} \;=\; \frac{L_m}{S_{m,i}}\!\bigl(R_m/R_i^{\max}\bigr)^{\gamma}.
\label{eq:instance-utilisation}
\end{aligned}
\end{equation}
The baseline $\,\alpha_i\,$ is the latency paid even at idle utilisation, whereas the second term $\beta_{m,i}\tilde\lambda_{m,i}^{\gamma}$ grows super-linearly once traffic increases.

\paragraph{\textbf{Empirical Validation.}}
Table~\ref{tab:inference_per_replica} reports the measured mean per-inference latencies of YOLOv5m ($m_2$) for different arrival rates $\lambda_{m_2}$ and replica counts $N_{m_2,i}$.
\begin{table}[h!]
\renewcommand{\arraystretch}{1.5}
\centering
\caption{The actual latency given by $\lambda_{m_2}=\{1, 2, 3, 4\}$ and $N_{m_2,i}=\{1, 2, 4\}$ per replica (e.g., $m_2$, YOLOv5m, 3 CPUs per replica) (seconds).}
\resizebox{0.5\textwidth}{!}{%
\begin{tabular}{|c|c|c|c|c|}
\hline
$N_{m,i}$ & $\lambda_m$ = 1 & $\lambda_m$ = 2 & $\lambda_m$ = 3 & $\lambda_m$ = 4\\
\hline
1 & 
$0.73{\scriptstyle\pm}0.004$ &
$4.97{\scriptstyle\pm}0.02$ &
$7.71{\scriptstyle\pm}0.03$ &
$10.46{\scriptstyle\pm}0.04$ \\
\hline
2 &
$0.73{\scriptstyle\pm}0.004$ &
$1.26{\scriptstyle\pm}0.19$ &
$3.76{\scriptstyle\pm}0.33$ &
$5.12{\scriptstyle\pm}0.53$ \\
\hline
4 & 
$0.73{\scriptstyle\pm}0.004$ &
$0.90{\scriptstyle\pm}0.06$ &
$1.12{\scriptstyle\pm}0.12$ &
$1.77{\scriptstyle\pm}0.29$ \\
\hline
\end{tabular}%
}
\label{tab:inference_per_replica}
\end{table}

\begin{figure}[ht]
    \centering
    \includegraphics[width=0.95\columnwidth]{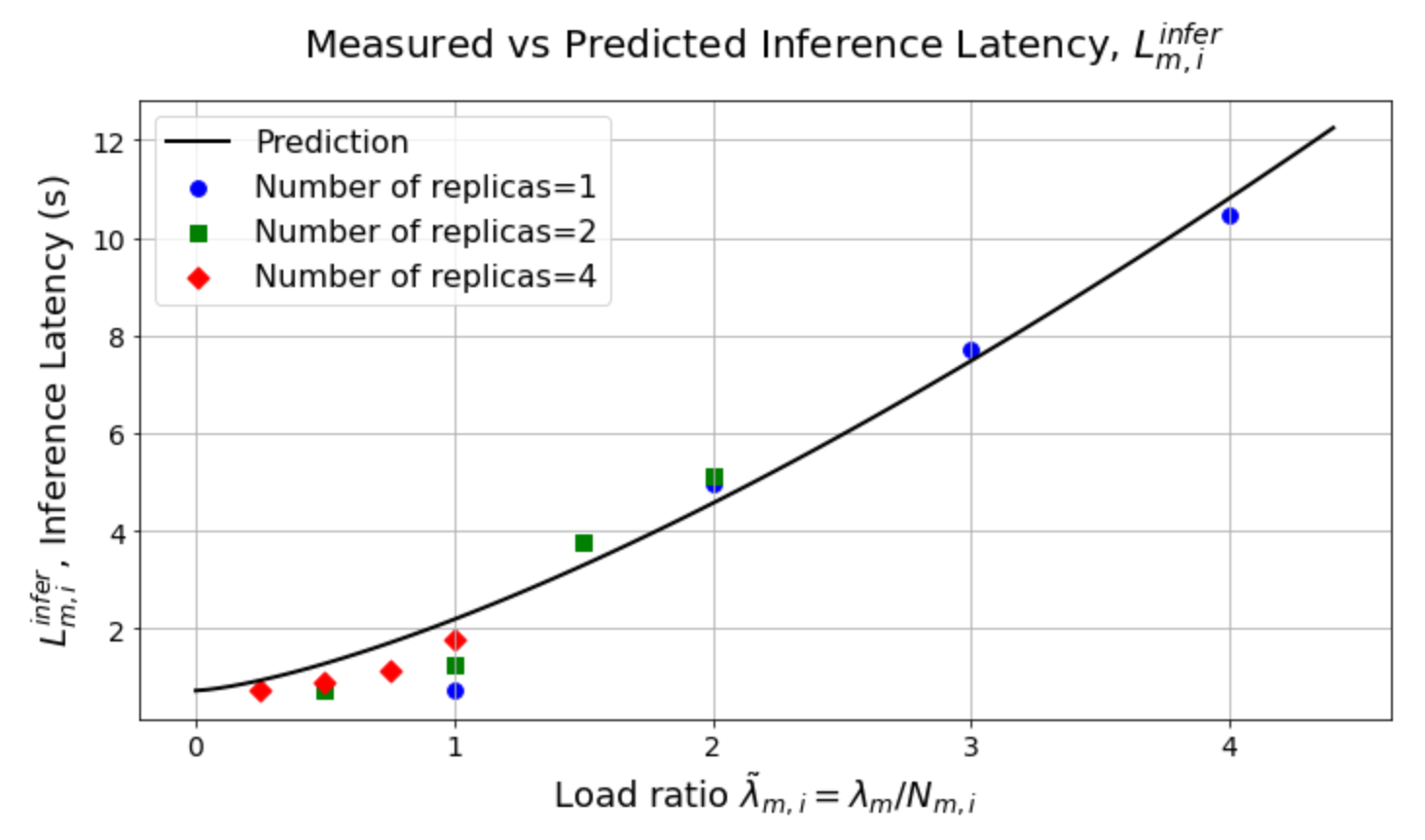}
    \caption{The inference latency measured in the real operations and predicted by Eq.~\ref{eq:inference-delay-final} with $\alpha_i$=0.73, $\beta_{m,i}$ = 1.29, and $\gamma$=1.49 (3 CPUs per replica).}
    \label{fig:inference_replica}
\end{figure}

Fig.~\ref{fig:inference_replica} shows that Eq.~\eqref{eq:inference-delay-final}, with calibrated parameters $\alpha_i\!=\!0.73$, $\beta_{m,i}\!=\!1.29$, and $\gamma\!=\!1.49$, closely matches the measurements.  Because the three parameters are re-estimated whenever the hardware mix ($S_{m,i},R_i^{\max}$) or co-tenant load ($B_i$) changes, the model remains accurate under a wide range of deployment conditions.  Such predictive capability is valuable for proactive resource provisioning and request routing during workload fluctuations.

\subsection{\textbf{Queueing Delay for Multi‑Replica Services}}
\label{subsec:queuing}
Let $N_{m,i}\!\in\!\mathbb{Z}_{>0}$ denote the \emph{replica count} of
model~$m$ on instance~$i$.
With exponential inter‑arrival and service times, the replica pool forms an $\text{M/M/}N_{m,i}$ queue.
The service rate is
\[
  \mu_{m,i}=S_{m,i}/L_m^{\text{infer}}
\]
and the traffic intensity
\[
  \rho_{m,i}=\lambda_m/(N_{m,i}\mu_{m,i}).
\]

\paragraph{Load Distribution.}  Tasks destined for \emph{model}~$m$ arrive at rate $\lambda_m$ and are distributed among the $N_{m,i}$ replicas via round‑robin or similar policies, yielding per‑replica arrival rate
\begin{equation}
  \lambda_{m,i}^{\text{replica}} = \frac{\lambda_m}{N_{m,i}}.
\end{equation}

\paragraph{M/M/$c$ Queue.}  Assuming exponential inter‑arrival and service times, each replica behaves as an M/M/$c$ queue with $c = N_{m,i}$ servers. Denoting the service rate by \(\mu_{m,i}=\tfrac{S_{m,i}}{L_m^{\text{infer}}},\) the traffic intensity becomes \(\rho_{m,i}=\tfrac{\lambda_m}{N_{m,i}\mu_{m,i}} .\)

Using Erlang‑$C$\,\cite{10.5555/1096491}
\begin{equation}
  C(\rho,c) \;=\;
  \frac{\dfrac{(\rho c)^c}{c!\,(1-\rho)}}
       {\displaystyle\sum_{k=0}^{c-1}\dfrac{(\rho c)^k}{k!}
        +\dfrac{(\rho c)^c}{c!\,(1-\rho)}},
  \label{eq:erlang‑c}
\end{equation}
the expected queueing delay becomes
\vspace{-4pt}
\begin{equation}
  Q_{m,i}(\lambda_m)
  \;=\;
  \frac{C\!\bigl(\rho_{m,i},N_{m,i}\bigr)}
       {N_{m,i}\mu_{m,i}-\lambda_m},
  \qquad
  \rho_{m,i}<1.
  \label{eq:queueing}
\end{equation}

A task bound to exactly one $(m,i)$ therefore experiences
\begin{equation}
  Q_{t,i}
  \;=\;
  \sum_{m\in\mathcal{M}}x_{t,m,i}\,
  Q_{m,i}(\lambda_m),
  \label{eq:task‑queue}
\end{equation}
which collapses to the unique non‑zero term selected by~$x_{t,m,i}$.

\subsection{\textbf{Task‑level Queue Delay Selection}}
Because each task is bound to exactly one \((m,i)\) via
\eqref{eq:assignment}, the queueing delay actually experienced by task
$t$ is the indicator‑weighted sum
\begin{equation}
  Q_{t,i} \;=\;
  \sum_{m\in\mathcal{M}} x_{t,m,i}\,Q_{m,i},
  \label{eq:q_task}
\end{equation}
which collapses to the unique non‑zero term for the chosen pair.

\begin{figure*}[ht]
    \centering

    \subfloat[Average Latency\label{fig:avg_latency}]{%
        \includegraphics[width=0.32\textwidth]{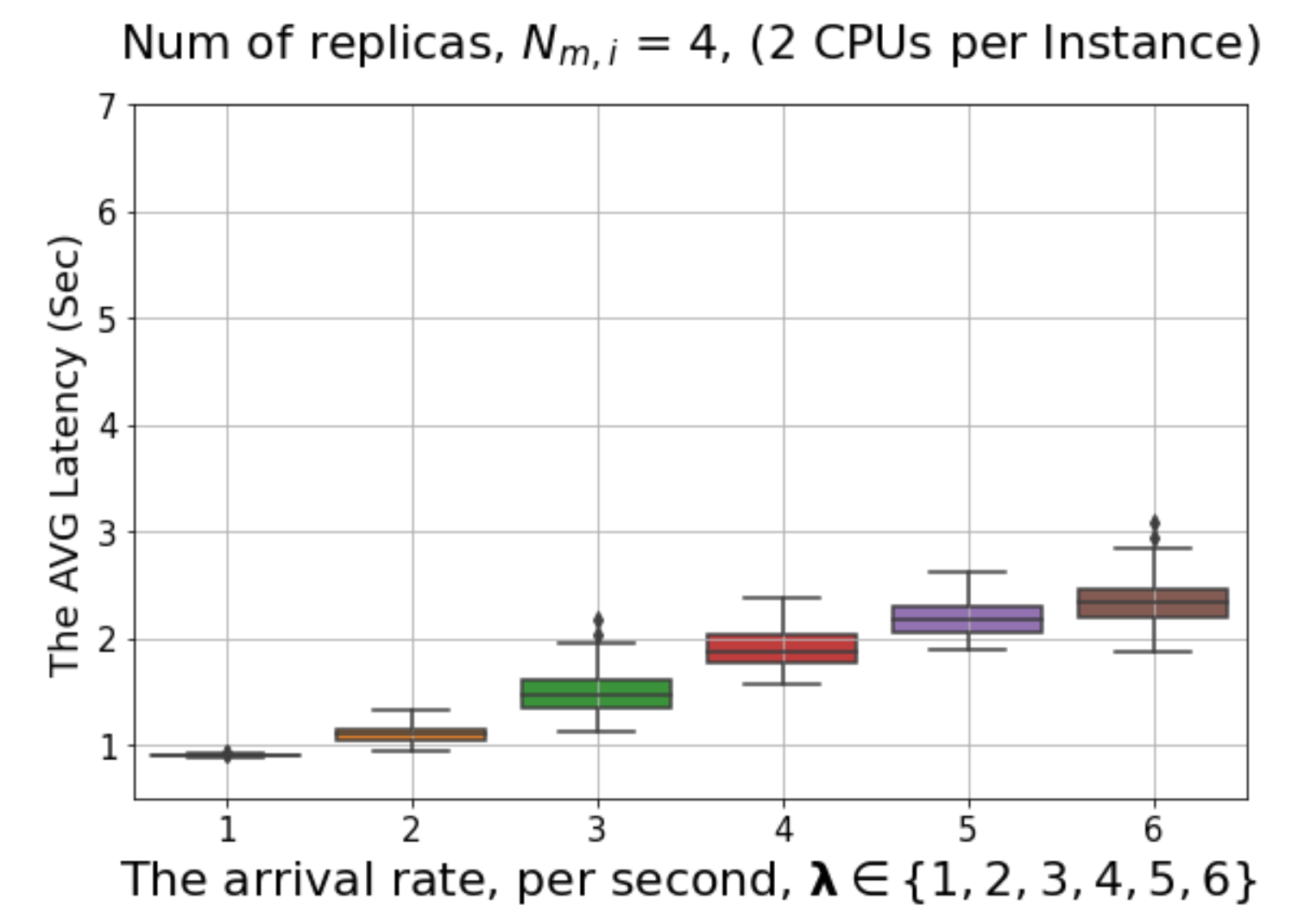}%
    }
    \hfill
    \subfloat[95th Percentile (P95) Latency\label{fig:p95_latency}]{%
        \includegraphics[width=0.32\textwidth]{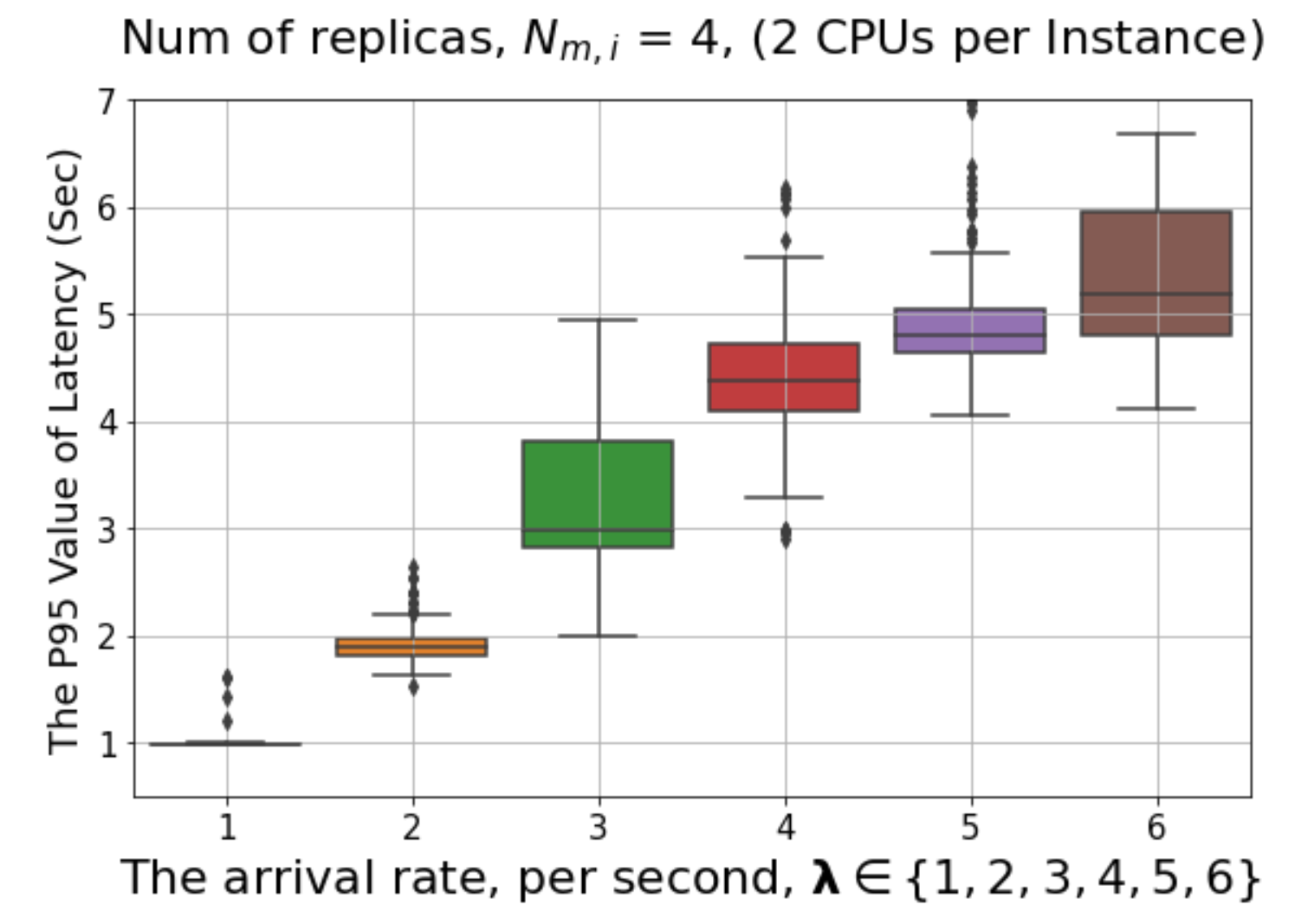}%
    }
    \hfill
    \subfloat[99th Percentile (P99) Latency\label{fig:p99_latency}]{%
        \includegraphics[width=0.32\textwidth]{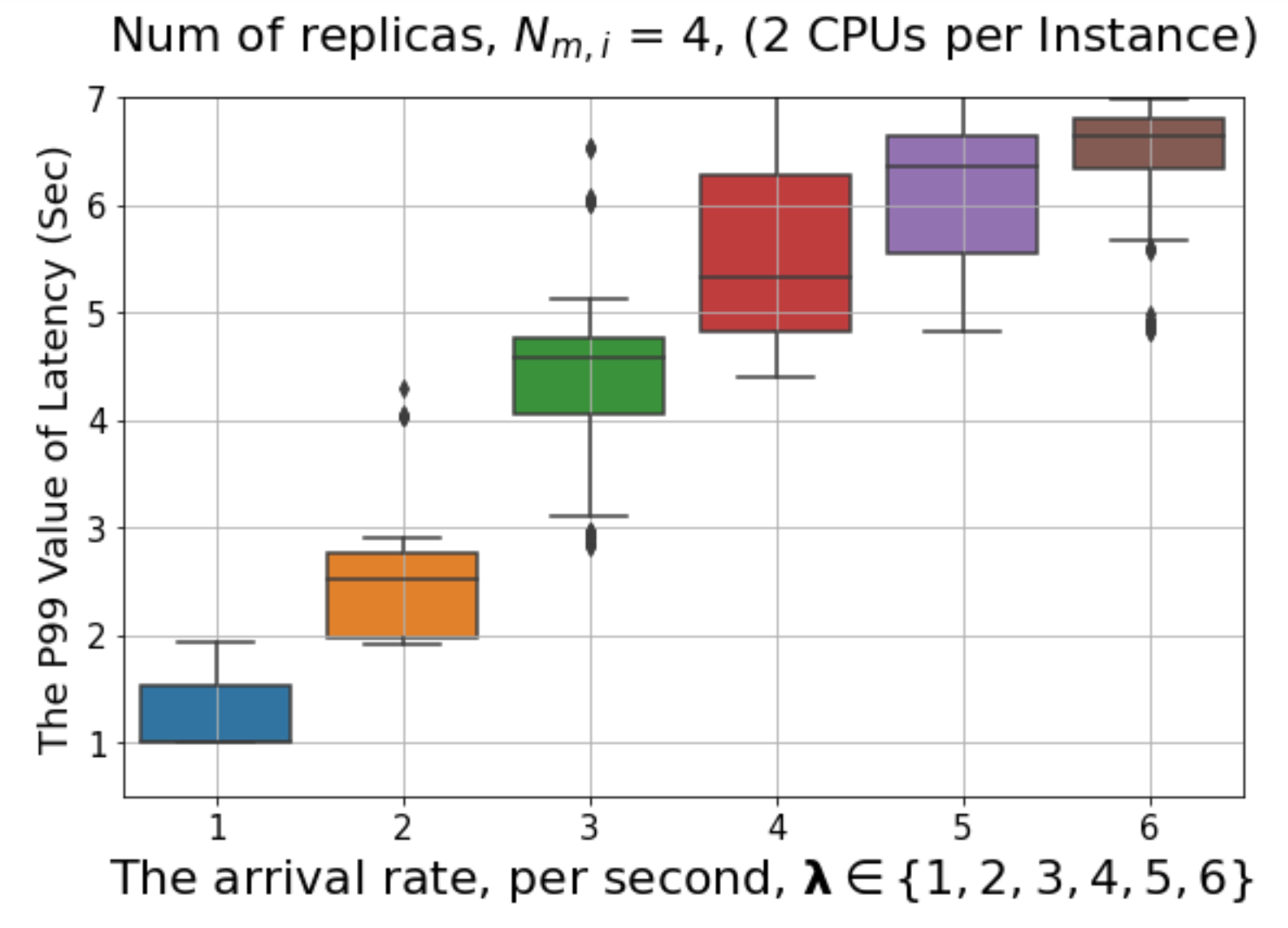}%
    }

    \caption{Latency metrics for user \texttt{robot19} under varying arrival rates, showing super-linear growth in average, P95, and P99 latencies.}
    \label{fig:latency_metrics}
\end{figure*}

\subsection{\textbf{Latency Function for Fixed Replica Layout}}
When the replica counts $\{N_{m,i}\}$ are \emph{fixed}, the
per‑instance end‑to‑end latency becomes an explicit function of the
arrival‑rate vector~$\boldsymbol\lambda$:
\begin{equation}
g_{m,i}(\boldsymbol\lambda)
\;=\;
\underbrace{\frac{L_m^{\text{infer}}}{S_{m,i}}
\Bigl[1+U_i(\boldsymbol\lambda)^{\gamma}\Bigr]}_{\text{processing}}
\;+\;
\underbrace{D_{m,i}^{\text{net}}}_{\text{network}}
\;+\;
\underbrace{\frac{C\!\bigl(\rho_{m,i},N_{m,i}\bigr)}{N_{m,i}\mu_{m,i}-\lambda_m}}_{\text{queueing}}
\label{eq:g‑function}
\end{equation}
with stability constraint $\rho_{m,i}\!<\!1$ for all $(m,i)$.

Substituting \eqref{eq:g‑function} into \eqref{eq:latency_components}
yields the task‑level latency
\begin{equation}
  L_t(\boldsymbol\lambda)
  \;=\;
  \sum_{m\in\mathcal{M}}\sum_{i\in\mathcal{I}}
  x_{t,m,i}\,
  g_{m,i}(\boldsymbol\lambda).
  \label{eq:latency‑final}
\end{equation}

Fig.~\ref{fig:latency_metrics} illustrates the service's latency characteristics across varying arrival rates $\boldsymbol\lambda = {1,\ldots,6}$ with $N_{m,i}=4$, revealing distinct behaviors for average, P95, and P99 latencies. The average latency increases gradually, reflecting growing queuing delays as load intensifies. In contrast, the P95 latency exhibits a steeper rise, indicating a broader spread in response times and the beginning of tail latency. The P99 latency escalates even more sharply, highlighting significant performance degradation under peak load conditions.

\subsection{\textbf{Per-Instance Latency as a Function of the Number of Replicas}}
\label{subsec:g_of_N}
With the arrival-rate vector $\boldsymbol\lambda$ held fixed, the only degree of freedom left in the per-instance latency expression is the replica count $N_{m,i}$.  Making this dependence explicit gives
\begin{equation}
\begin{aligned}
g_{m,i}\!\bigl(N_{m,i}\bigr)
&\;=\;
\underbrace{\frac{L_m^{\text{infer}}}{S_{m,i}}
\Bigl[1+U_i(\boldsymbol\lambda)^{\gamma}\Bigr]}_{\text{processing (constant)}}\;+\; \underbrace{D_{m,i}^{\text{net}}}_{\text{network (constant)}}\; \\
&+\; \underbrace{\frac{C\!\bigl(\rho_{m,i}(N_{m,i}),\,N_{m,i}\bigr)}{N_{m,i}\mu_{m,i}-\lambda_m}}_{\text{queueing (varies with }N_{m,i}\text{)}},
\label{eq:g-of-N}
\end{aligned}
\end{equation}
\noindent
where
\[
  \rho_{m,i}\!\bigl(N_{m,i}\bigr)
  \;=\;
  \frac{\lambda_m}{N_{m,i}\mu_{m,i}}
  \;<\;1
  \quad
  \text{(stability constraint).}
\]

Processing and network delays are unaffected by the replica count once $\boldsymbol\lambda$ is fixed. Queueing delay shrinks as $N_{m,i}$ grows because both the service-pool capacity $N_{m,i}\mu_{m,i}$ increases linearly and the utilisation $\rho_{m,i}(N_{m,i})$ falls hyperbolically.

As an $\mathrm{M/M/}N_{m,i}$ model is required, we adjust the factor $C(\rho,N)$ by using the Erlang-$C$ formula:
\[C(\rho,N)=
\frac{\dfrac{\rho^{N}}{N!\,(1-\rho)}}
     {\displaystyle
      \sum_{k=0}^{N-1}\frac{\rho^{k}}{k!} \;+\;
      \frac{\rho^{N}}{N!\,(1-\rho)}
     },
\qquad
\rho=\frac{\lambda_m}{N\mu_{m,i}}.
\]

The marginal benefit of adding replicas is largest near the
instability boundary $(N_{m,i}\mu_{m,i}\gtrsim\lambda_m)$ and flattens rapidly once $\rho_{m,i}\!\lesssim\!0.3$.  This shape is crucial for choosing a cost-optimal replica layout that still satisfies latency SLOs.

Therefore, we define the task-level latency by substituting \eqref{eq:g-of-N} into
\[
  L_t(\{N_{m,i}\})
  \;=\;
  \sum_{m\in\mathcal{M}}\sum_{i\in\mathcal{I}}
    x_{t,m,i}\,g_{m,i}(N_{m,i})
\]
yields a closed-form, differentiable objective that can be handed for automatic replica-layout tuning.

\subsection{\textbf{Optimisation Problems with Two Latency Models}} \label{subsec:optimisation}

The latency of a request routed through replica group $(m,i)$ can be expressed in \emph{two} complementary closed‐form models:
\[
  \underbrace{g_{m,i}(\boldsymbol\lambda)}_{\text{fixed }N_{m,i}}
  \quad\text{and}\quad
  \underbrace{g_{m,i}\!\bigl(N_{m,i}\bigr)}_{\text{fixed }\boldsymbol\lambda}.
\]

Define the corresponding task-level latencies
\[
  L_t^{(\lambda)}
  :=\sum_{m,i}x_{t,m,i}\,g_{m,i}(\boldsymbol\lambda),
  \qquad
  L_t^{(N)}
  :=\sum_{m,i}x_{t,m,i}\,g_{m,i}(N_{m,i}).
\]

Two optimisation stages naturally arise.

\paragraph{Workload routing (\emph{fixed replica layout, $N_{m,i}$)}} It is about a problem to route which replica is choosen out of the $N_{m,i}$ replicas. Given the current replica counts $\{N_{m,i}\}$ and arrival-rate vector
$\boldsymbol\lambda$, the router chooses $x$:

\begin{align}
  \min_{x}\;&\;
     \max_{t\in\mathcal{T}} L_t^{(\lambda)}
     \label{eq:stageA-obj}\\[2pt]
  \text{s.t.}\quad
  &\sum_{m,i}x_{t,m,i}=1,&\forall t\in\mathcal{T}
     \label{eq:stageA-route}\\
  &\sum_{t,m}x_{t,m,i} R_m \le R_i^{\max},&\forall i\in\mathcal{I}
     \label{eq:stageA-cap}\\
  &L_t^{(\lambda)}\le \tau_t,&\forall t\in\mathcal{T}
     \label{eq:stageA-slo}\\
  &\rho_{m,i}(\lambda_m)<1,&\forall m,i\;(\text{stability}).
     \label{eq:stageA-stab}
\end{align}

\paragraph{Capacity planning \& routing (\emph{fixed traffic})}
For longer-term provisioning the operator sizes the replica pools and
chooses routing simultaneously:

\begin{align}
  \min_{\{N_{m,i}\},x}\;&\;
     \max_{t\in\mathcal{T}} L_t^{(N)}
     \;+\;
     \beta\!\sum_{m,i} c_{m,i}\,N_{m,i}
     \label{eq:stageB-obj}\\[2pt]
  \text{s.t.}\quad
  &\eqref{eq:stageA-route}--\eqref{eq:stageA-cap},\;
    L_t^{(N)}\le\tau_t\ \forall t
     \label{eq:stageB-slo}\\
  &\lambda_m < N_{m,i}\mu_{m,i},\;\;\forall m,i
     \label{eq:stageB-stab}\\
  &N_{m,i}\in\mathbb{Z}_{\ge 1},\;\;\forall m,i.
     \label{eq:stageB-int}
\end{align}

\smallskip
\noindent
Here $c_{m,i}$ is the per-replica cost and $\beta$ trades off latency
versus spend.

\section{Latency-Aware, Predictive In-Memory Routing and Proactive Autoscaling (LA-IMR)}\label{sec:4}

Based on the sub-linear model proposed in~\S\ref{subsec:processing_delay} and the instance utilisation-oriented delay expression in~\eqref{eq:g‑function} and~\eqref{eq:g-of-N}, We design a control layer that \emph{schedules, routes, and offloads} requests so that \emph{tail-latency} (\(P99\)) stays within each task’s SLO~\(\tau_t\) even under bursty traffic and heterogeneous hardware.

Furthermore, the proposed LA-IMR framework leverages the modular nature of microservice architecture to improve overall service quality. By reducing tail latency—the slowest response times—it ensures more consistent and predictable system performance.

\medskip
\noindent LA-IMR comprises three tightly-coupled components:

\begin{figure*}[ht]
\centering

\subfloat[Average Latency\label{fig:avg_comparision_micro_monolithic}]{%
    \includegraphics[width=0.32\textwidth]{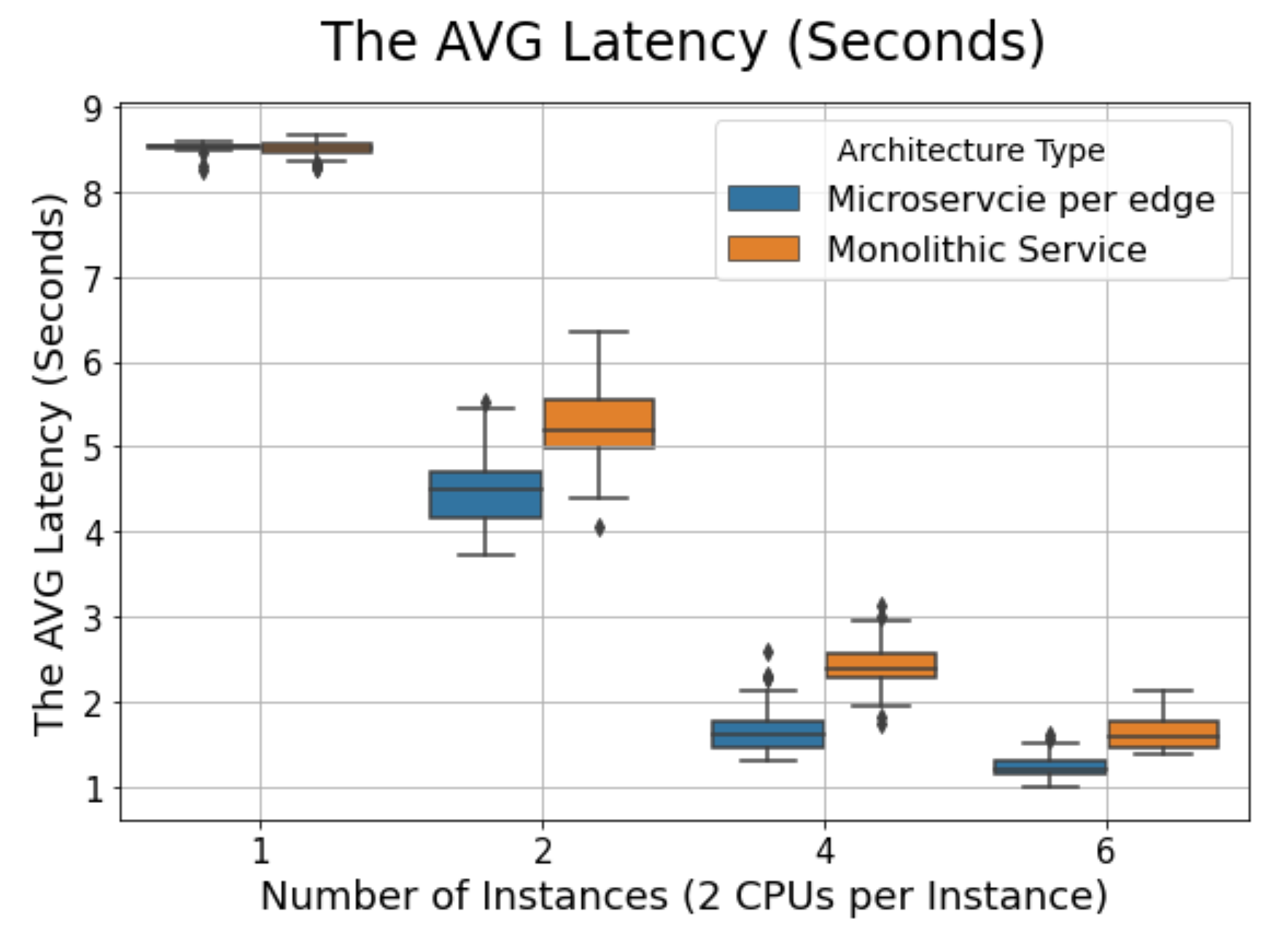}%
}
\hfill
\subfloat[95th Percentile (P95) Latency\label{fig:p95_comparision_micro_monolithic}]{%
    \includegraphics[width=0.32\textwidth]{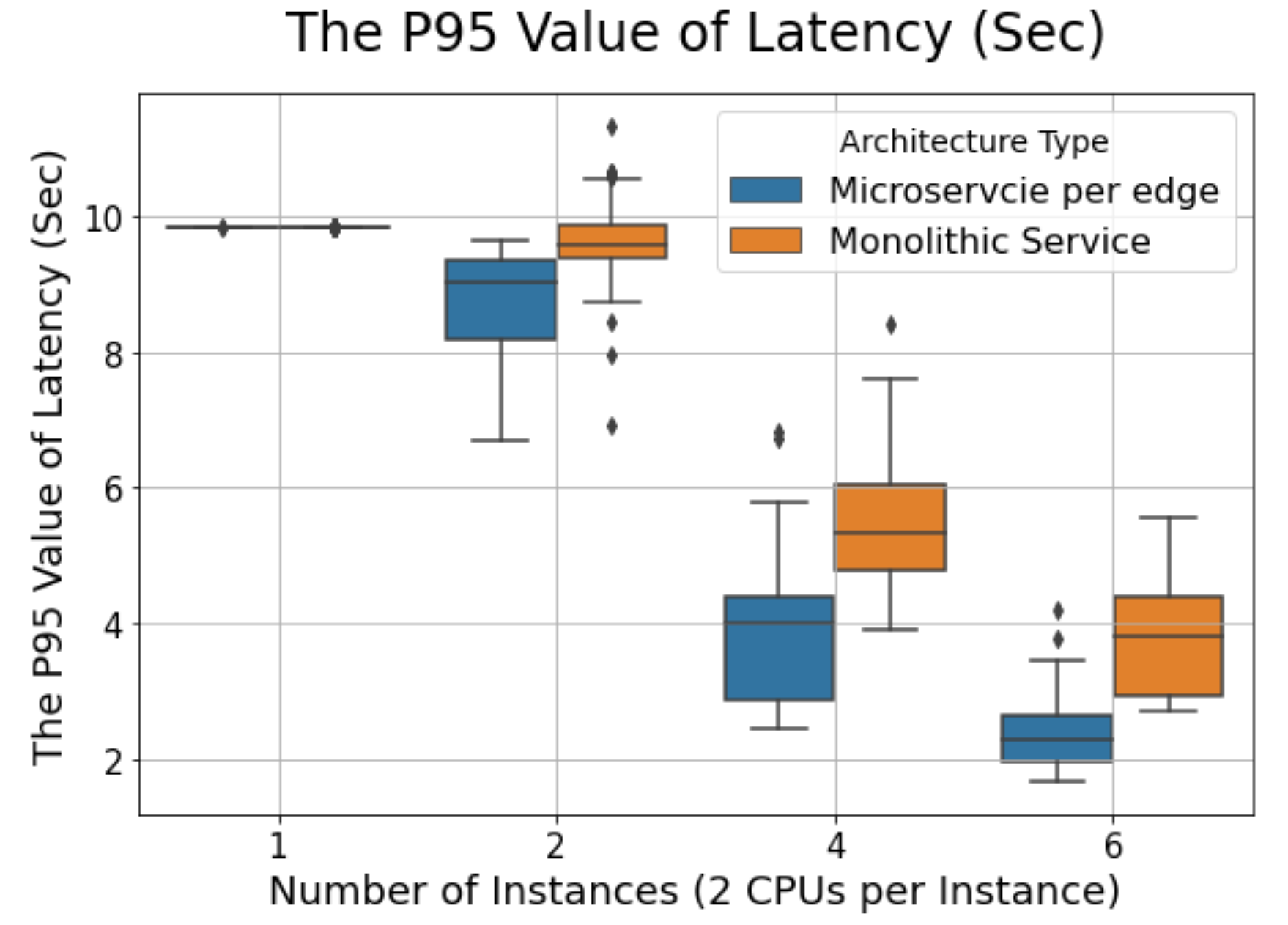}%
}
\hfill
\subfloat[99th Percentile (P99) Latency\label{fig:p99_comparision_micro_monolithic}]{%
    \includegraphics[width=0.32\textwidth]{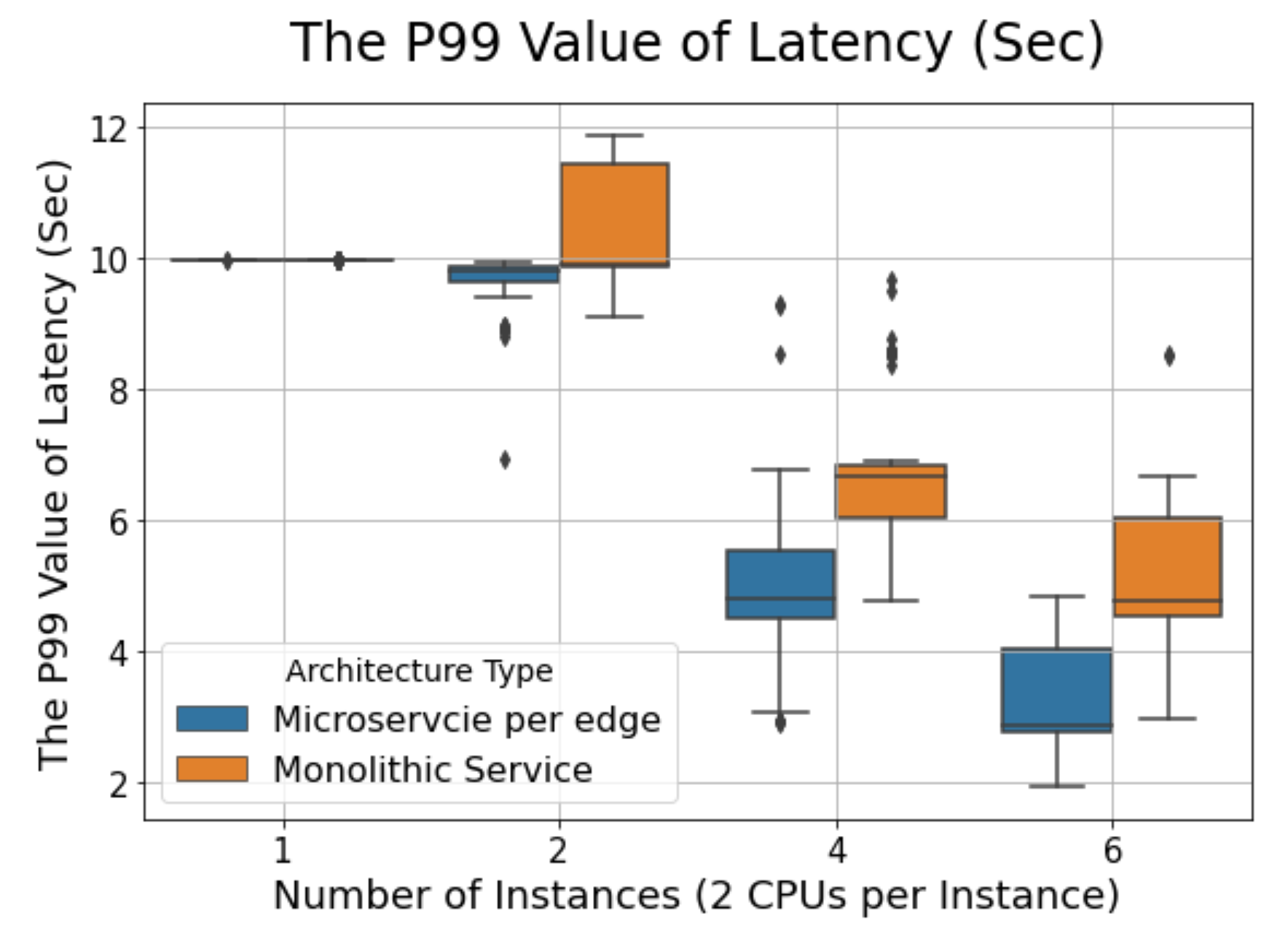}%
}

\caption{Inference latency comparison between the microservice and monolithic service architecture as the number of replica $N_{m,i}$ increases when the arrival rate $\lambda$=4 is given. Overall, the microservice architecture shows the superior latency.}
\label{fig:comparision_micro_monolithic}
\end{figure*}

\subsection{\textbf{Quality-Differentiated Multi-Queue Scheduler}}
\label{subsec:LAIMR-queues}
The SLO-Aware Inference Router maintains and monitors the status of the different SLO requests by using its corresponding queue at the code level, leading to the real-time monitoring and the early-latency spiks detection.

To address diverse quality of service (QoS) requirements -- such as accuracy and latency -- we decompose inference capabilities into specialized \emph{microservices} aligned with following performance tiers:
\begin{itemize}
    \item Low-Latency Services (e.g., edge-optimized service like ultra-low latency): Lightweight models such as EfficientDet are deployed on resource-constrained edge nodes to support real-time, latency-sensitive tasks.
    \item Balanced Services: Mid-range models such as YOLOv5m offer a trade-off between latency and accuracy, ideal for tasks with moderate performance demands.
    \item Precision Services (e.g., accuracy-prioritized): Computationally heavier models such as Faster R-CNN run in the cloud, delivering  high accuracy for use cases where latency is less critical.
\end{itemize}

We partition traffic into \emph{quality classes} \(\mathcal{Q}=\{\texttt{Low-Latency},\texttt{BALANCED}, \texttt{PRECISE}\)\}, each backed by an run-time queue \(\mathcal{Q}_q\).

\begin{itemize}[leftmargin=1.2em]
  \item Low-Latency lane (\texttt{Low-Latency})\,: latency-critical tasks use small-footprint EfficientDet-Lite0 streams and inherit the highest dispatch priority.
  \item BALANCED lane (\texttt{BALANCED})\,:  accuracy-bound tasks are serviced by YOLO5m replicas and accept longer—but still bounded—delays.
  \item PRECISE lane (\texttt{PRECISE})\,:  accuracy-bound tasks (e.g.\ fine-grained inspection) are serviced by R-CNN replicas.
\end{itemize}
\begin{table}[h]
\centering
\caption{Comparison of YOLOv5m and EfficientDet-Lite0}
\begin{tabular}{|l|c|c|}
\hline
\textbf{Attribute} & \textbf{YOLOv5m}~\cite{glenn_jocher_2020_4154370} & \textbf{EfficientDet-Lite0}~\cite{tan2020efficientdet} \\
\hline
Architecture        & Ultralytics YOLOv5       & Google EfficientDet-Lite \\
Model Size          & 21.2M                    & 4.3M                     \\
mAP@0.5             & 64.1\%                   & \textasciitilde \ 25\%     \\
mAP@0.5:0.95        & 45.4\%                   & \textasciitilde \ 20\%     \\
Use Case            & Balanced performance     & Edge/mobile efficiency   \\
\hline
\end{tabular}
\end{table}

Due to its high modularity, the microservice architecture offers several advantages over the monolithic service architecture, where all necessary services are deployed within a single instance. Fig.\ref{fig:comparision_micro_monolithic} presents a comparison of the latency between the two architectures. Fig.\ref{fig:avg_comparision_micro_monolithic} shows the average latency, Fig.\ref{fig:p95_comparision_micro_monolithic} illustrates the 95th percentile (P95) latency, and Fig.\ref{fig:p99_comparision_micro_monolithic} presents the 99th percentile (P99) latency for both architectures. Overall, the microservice architecture demonstrates superior latency performance—especially when sufficient resources, such as an increased number of replicas $N_{m,i}$, are available. This is primarily because context switching among different models imposes a higher burden on a monolithic-based instance.

\subsection{\textbf{SLO-Aware Adaptive Routing}}\label{subsec:LAIMR-routing}

A dedicated \emph{SLO‑Aware Inference Router} (Fig.~\ref{fig:architecture}) dynamically forwards robotic and general inference requests to the most suitable micro‑service tier.  
Each arriving request is represented as the tuple
\[
  r=\bigl(m,i,t\bigr),\qquad
  \lambda_m=\textsc{SlidingRate}(m,t)
\]
where \(m\) is the model, \(i\) the tier index, \(t\) the arrival time, and \(\lambda_m\![\mathrm{req/s}]\) is the 1‑s sliding‑window arrival rate maintained in memory.

\begin{figure}[ht]
    \centering
    \includegraphics[width=0.95\columnwidth]{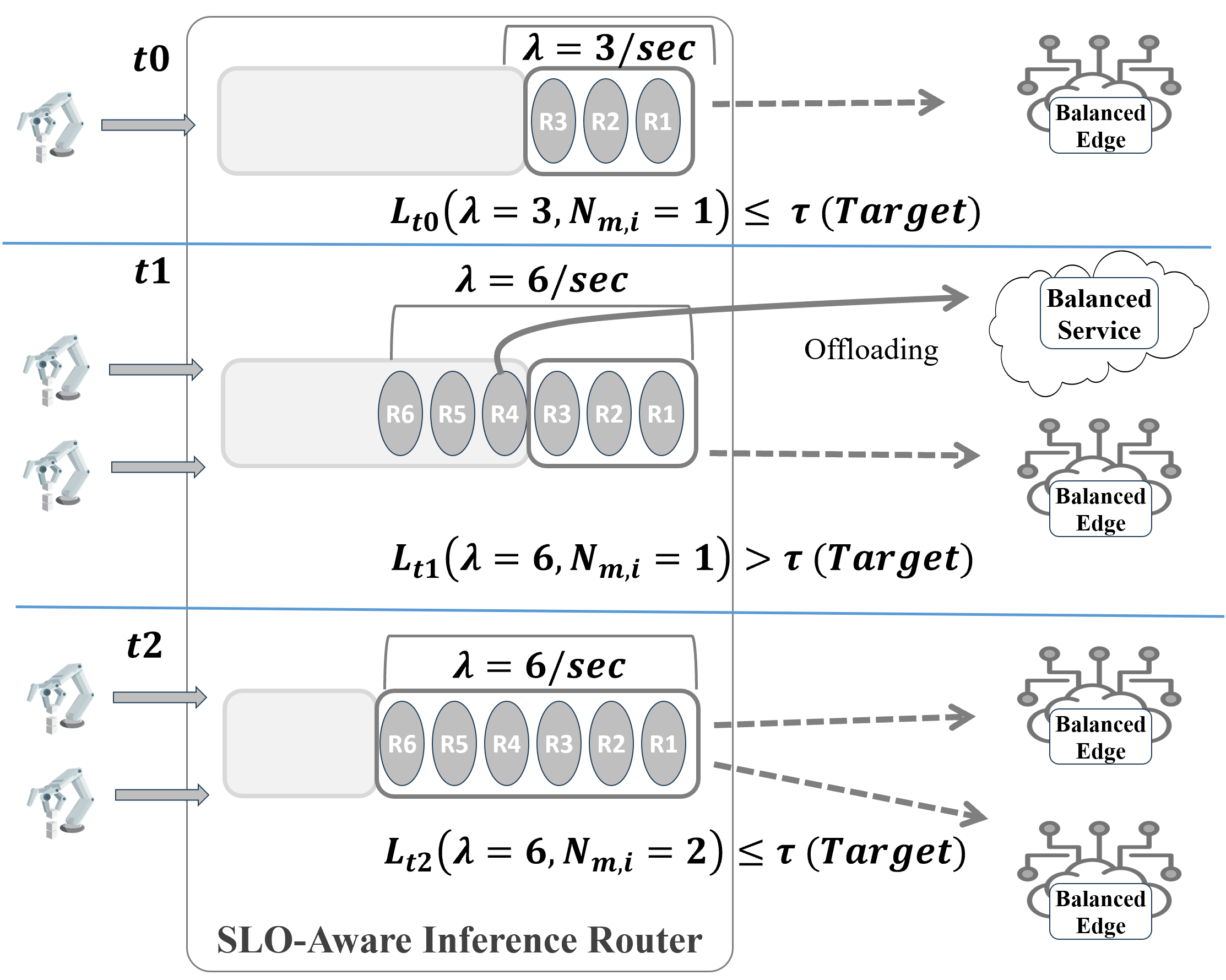}
    \caption{Real-time latency prediction uses the arrival rate $\lambda$ to meet the target latency $\tau$. If latency exceeds $\tau$, the system increases replicas $N_{m,i}$. This prediction also enables proactive offloading based on $\lambda$ and $N_{m,i}$.}
    \label{fig:SLO_Aware_inference_router}
\end{figure}

Given the current replica layout \(\{N_{m,i}\}\) and the instantaneous arrival‑rate vector \(\boldsymbol\lambda\), the router executes the following steps:

\begin{enumerate}[leftmargin=1.2em,label=\roman*)]
\item \textbf{Compute the model‑specific latency budget}:  
      \(\displaystyle
        \tau_m
        \;=\;
        x\,L_m^{\text{infer}}
      \)
      with a global multiplier \(x{>}1\) that budgets headroom for networking and queueing delays shown in Algorithm~\ref{alg:laimr:xscaled}.

\item \textbf{Predict per‑instance latency}:  
      look up \(g_{m,i}(\boldsymbol\lambda)\) in an in‑memory table pre‑computed by the analytic model  
      (\S\ref{subsec:processing_delay},\ref{subsec:g_of_N}) and refreshed every \(\Delta\)~seconds.

\item \textbf{Filter feasible replicas}:  
      retain only the pairs \(\langle m,i\rangle\) whose predicted latency satisfies the SLO  
      \(g_{m,i}(\boldsymbol\lambda)\le\tau_m\).

\item \textbf{Select the target replica}:  
      choose  
      \[
        (m^\star,i^\star)
        \;=\;
        \arg\min_{\langle m,i\rangle}
        g_{m,i}(\boldsymbol\lambda),
      \]
      breaking ties by the lower cost \(c_{m,i}\) to avoid unnecessary over‑provisioning.

\item \textbf{Enqueue the request}:  
      push \(r\) into the queue of tier~\(i^\star\).  
      If no local replica meets the budget (\(g_{m,i}>\tau_m\;\forall i\)), offload \(r\) to the upstream (faster or cloud) tier as prescribed by Algorithm~\ref{alg:laimr:xscaled}.
\end{enumerate}

\subsection{\textbf{Edge–Cloud Offloading \& Replica Autoscaling}}\label{subsec:LAIMR-offload}
The SLO-Aware Inference Router detects the workload fluctuation and decide when to offload to the cloud and other edge service by using the per-instance latency determined by the inputs of the arrval rate and the current resource, such as the number of the replicas.

The event-driven LA-IMR controller reacts on every incoming request instead of running at fixed intervals. It first computes a 1-second sliding-window arrival rate $\lambda_m$ to judge whether the just-arrived request would breach the latency SLO; if so, the request is immediately off-loaded to the faster/cloud tier. In parallel, it maintains an EWMA-smoothed accumulated rate $\lambda_m^{accumul}$ that captures sustained demand. This second metric drives replica scaling and bulk off-load decisions, ensuring resources grow only when higher load persists and shrink when utilisation stays low shown in Fig.~\ref{fig:SLO_Aware_inference_router}. By combining per-request mitigation with stable long-term control, the algorithm keeps tail-latency low while avoiding unnecessary resource oscillations, detailed in Algorithm~\ref{alg:laimr:xscaled}.

\begin{algorithm}[t]
\SetAlgoLined
\KwIn{incoming request $r$ for service instance $(m,i)$ at time $t_\text{now}$}
\SetKwProg{Fn}{Function}{:}{}
\Fn{$\textsc{SlidingRate}(m,\;t_\text{now})$}{
    \While{$Q_m\neq\varnothing$ \textbf{and} $t_\text{now}-Q_m.\text{front}() > 1$}{
        $Q_m.\text{pop\_front}()$ \tcp*{discard arrivals $>$1 s old}
    }
    $Q_m.\text{push\_back}(t_\text{now})$\;
    \Return{$\lambda_m \leftarrow |Q_m| \;[\mathrm{req/s}]$}\;
}

\SetKwInput{KwParam}{\textbf{Parameters}}
\KwParam{$x>1$ (latency multiplier), EWMA weight $\alpha$, utilization floor $\rho_{\text{low}}$, per-instance replica cap $N_{m,i}^{\max}$}

$\lambda_m  \leftarrow \textsc{SlidingRate}(m,\,t_\text{now})$\;
$\tau_m  \leftarrow x\,L_m^{\text{infer}}$    \tcp*{\bf Per-model SLO}
$\widehat{g}^{\text{inst}}_{m,i} \leftarrow g_{m,i}(\lambda_m)$\;

\If(\tcp*[f]{protect this single request}){$\widehat{g}^{\text{inst}}_{m,i} > \tau_m$}{
    \textbf{offload} $r$ to nearest fast/cloud tier\;
    \Return
}

read $N_{m,i},\;\rho_{m,i}$ from shared state\;
$\lambda_m^{\text{accum}} \leftarrow
        \alpha\,\lambda_m^{\text{accum}} + (1-\alpha)\,\lambda_m$\;
$\widehat{g}_{m,i} \leftarrow g_{m,i}(\lambda_m^{\text{accum}})$\;

\If(\tcp*[f]{predicted SLO breach}){$\widehat{g}_{m,i} > \tau_m$}{
    \eIf{$N_{m,i} < N_{m,i}^{\max}$}{
        \textbf{scale\_out} one replica on the current tier\;
    }{
        $\displaystyle
        \phi \leftarrow
           \min\!\Bigl(1,\,
                 \frac{\widehat{g}_{m,i}-\tau_m}{\widehat{g}_{m,i}}
               \Bigr)$\;
        \textbf{offload} fraction $\phi$ upstream
        (balanced\,$\!\rightarrow\!$\,low-latency tier)\;
    }
}
\ElseIf{$\rho_{m,i} < \rho_{\text{low}}$ \textbf{and} $N_{m,i} > 1$}{
    \textbf{scale\_in} one replica to save cost\;
}

route request $r$ to the chosen local replica\;
\caption{Event-driven LA-IMR with $x$-scaled latency SLO}
\label{alg:laimr:xscaled}
\end{algorithm}

\subsection{\textbf{Implementation of Dynamic Scaling in a Kubernetes Environment}} \label{subsec:dynamic_scaling}

We realise the replica decisions of Algorithm~\ref{alg:laimr:xscaled} with the
\emph{Kubernetes Horizontal Pod Autoscaler~(HPA)}.
Instead of relying on generic resource indicators such as~CPU~\%,
the LA‑IMR controller exports a \texttt{desired\_replicas} \textbf{custom metric} for every Deployment \(\langle m,i\rangle\):
\[
  \texttt{desired\_replicas}_{m,i}
  \;=\;
  N_{m,i}(t),
\]
where \(N_{m,i}(t)\) is the replica count computed in line~15 of Algorithm~\ref{alg:laimr:xscaled}.

The metric is scraped by Prometheus and surfaced to the HPA through the \texttt{k8s‑prometheus‑adapter}. The HPA’s reconciliation loop then executes, every~5 s:
\begin{enumerate}[leftmargin=1.2em,label=\roman*)]
    \item \textbf{Read the custom metric} and compare it with the current Pod count.
    \item \textbf{Scale out} (or in) by the exact difference, bounded by the per‑Deployment cap \(N_{m,i}^{\max}\) and cluster quotas.
    \item \textbf{Respect graceful‑termination}: drained Pods are held until in‑flight requests finish, preventing mid‑request losses.
\end{enumerate}

Because the scaling trigger is the \emph{predicted} latency budget (\(\tau_m = x\,L_m^{\text{infer}}\)) rather than lagging utilisation, extra replicas are spun up \textit{before} queueing delay violates the SLO and are shed once utilisation drops below \(\rho_{\text{low}}\). In practice this removes the 60–120 s reaction lag typical of threshold‑based autoscalers, keeps the \(p_{99}\) latency inside the \(x\,L_m^{\text{infer}}\) envelope, and avoids chronic over‑provisioning.

\section{Performance Analysis}  \label{sec:5}

\subsection{\textbf{Experiment Environment}}

\subsubsection{\textbf{Hardware Specifications}}
\begin{figure}[htbp]
    \centering

    \subfloat[Cube manipulation.\label{fig:image1}]{%
        \includegraphics[width=0.48\columnwidth]{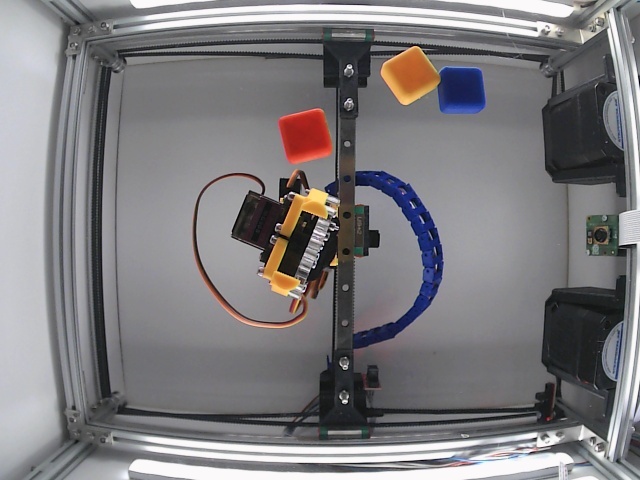}%
    }
    \hfill
    \subfloat[Strip manipulation.\label{fig:image2}]{%
        \includegraphics[width=0.48\columnwidth]{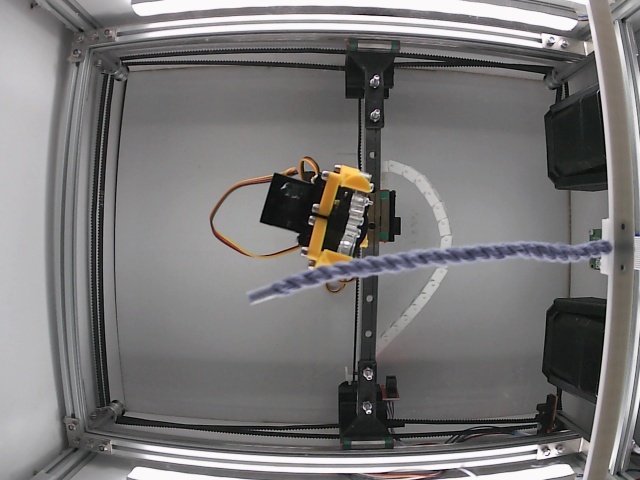}%
    }

    \caption{CloudGripper work cells performing object manipulation with dual-camera views and consistent cell configurations.}
    \label{fig:two_images}
\end{figure}

We utilize the CloudGripper testbed—a scalable, open-source, rack-mounted cloud robotics platform optimized for large-scale manipulation tasks. Each cell features a low-cost, 5-DOF Cartesian robot with a rotatable parallel-jaw gripper, controlled via a Raspberry Pi 4B (Quad-core Cortex-A72, 1.8GHz) and a Teensy 4.1 for real-time actuation. Dual RGB cameras (top and bottom views) capture multi-angle data at up to 30 FPS. As shown in Fig.~\ref{fig:two_images}, each robot operates in a standardized, enclosed cell (274 mm × 356 mm × 400 mm) with dedicated lighting, ensuring uniform and parallelized data collection. Five of them are connected to our SLO-Aware interference router to get services by sending an image taken by its camera and receiving the coordinates of the object.

\subsubsection{\textbf{Cloud/edge Computing Continuum}}
CloudGripper leverages a cloud-edge continuum to support scalable, distributed robotic control and experimentation. The edge infrastructure includes a 32-robot rack connected via 1 Gbit/s Ethernet to a 10 Gbit/s switch, forming an on-campus edge cluster. Each robot is operated through a REST API on a Raspberry Pi 4B, enabling low-latency control and dual-camera streaming at 30 FPS. The edge cluster comprises 32 Raspberry Pis (32x4 cores) running on Kubernetes with Prometheus-based monitoring, achieving an average container startup time of 1.8 seconds on ARM64.

A remote cloud cluster, hosted by Ericsson, supplements this with 19 dedicated CPU cores and a 36 ms network delay over a 10 Gbit/s link.

This setup enables dynamic offloading between edge and cloud, optimizing for latency, resource constraints, and performance. Benchmarks highlight the impact of deployment location on responsiveness, underscoring the need for adaptive orchestration in real-time robotics.

\begin{figure*}[ht]
\centering

\subfloat[Average, P95, and P99 latencies of the proposed LA-IMR approach.
\label{fig:comparison_la_imr}]{%
    \includegraphics[width=0.48\textwidth]{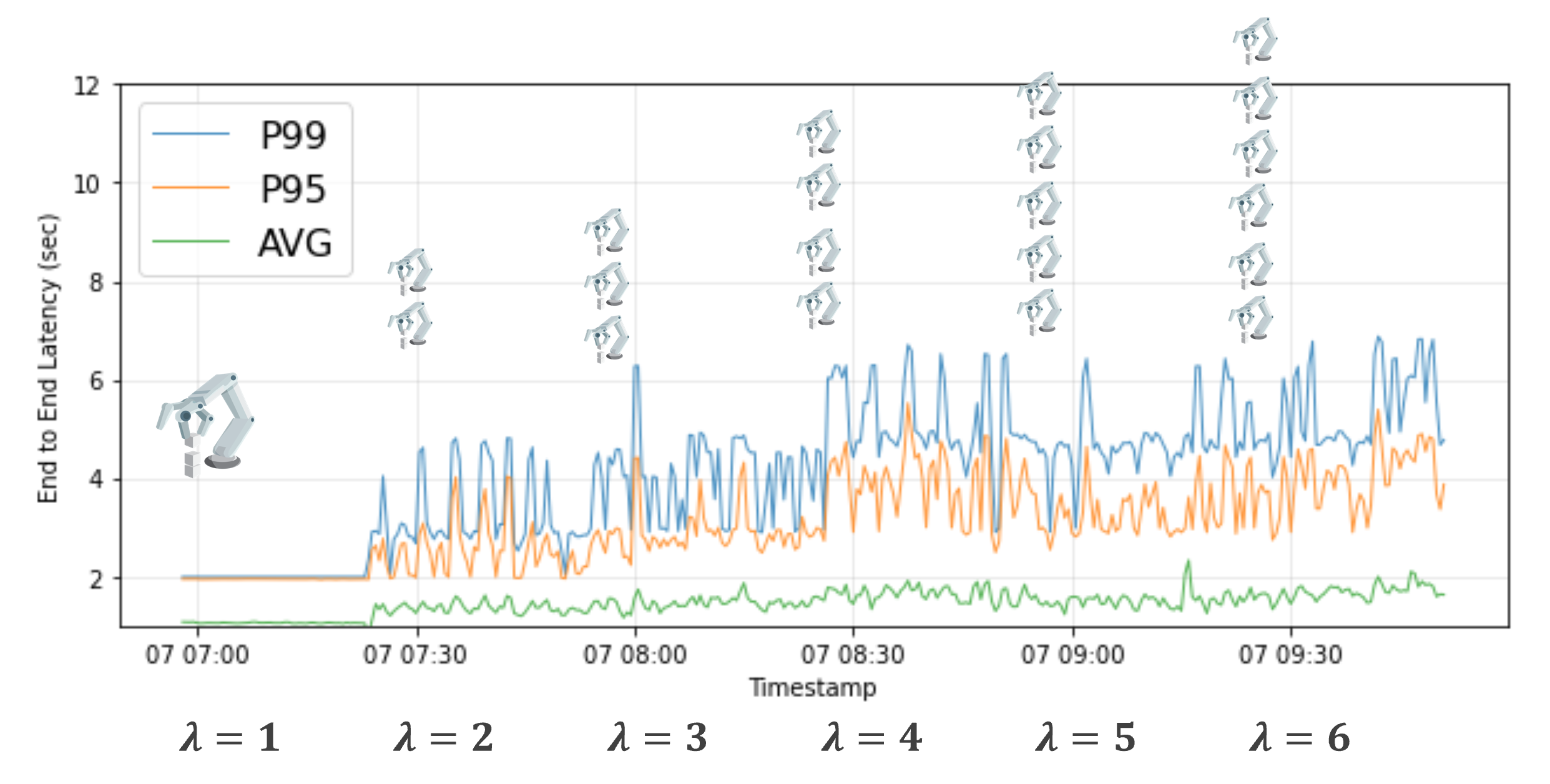}%
}
\hfill
\subfloat[The latencies of the baseline method using Prometheus-measured latency.
\label{fig:comparison_baseline}]{%
    \includegraphics[width=0.48\textwidth]{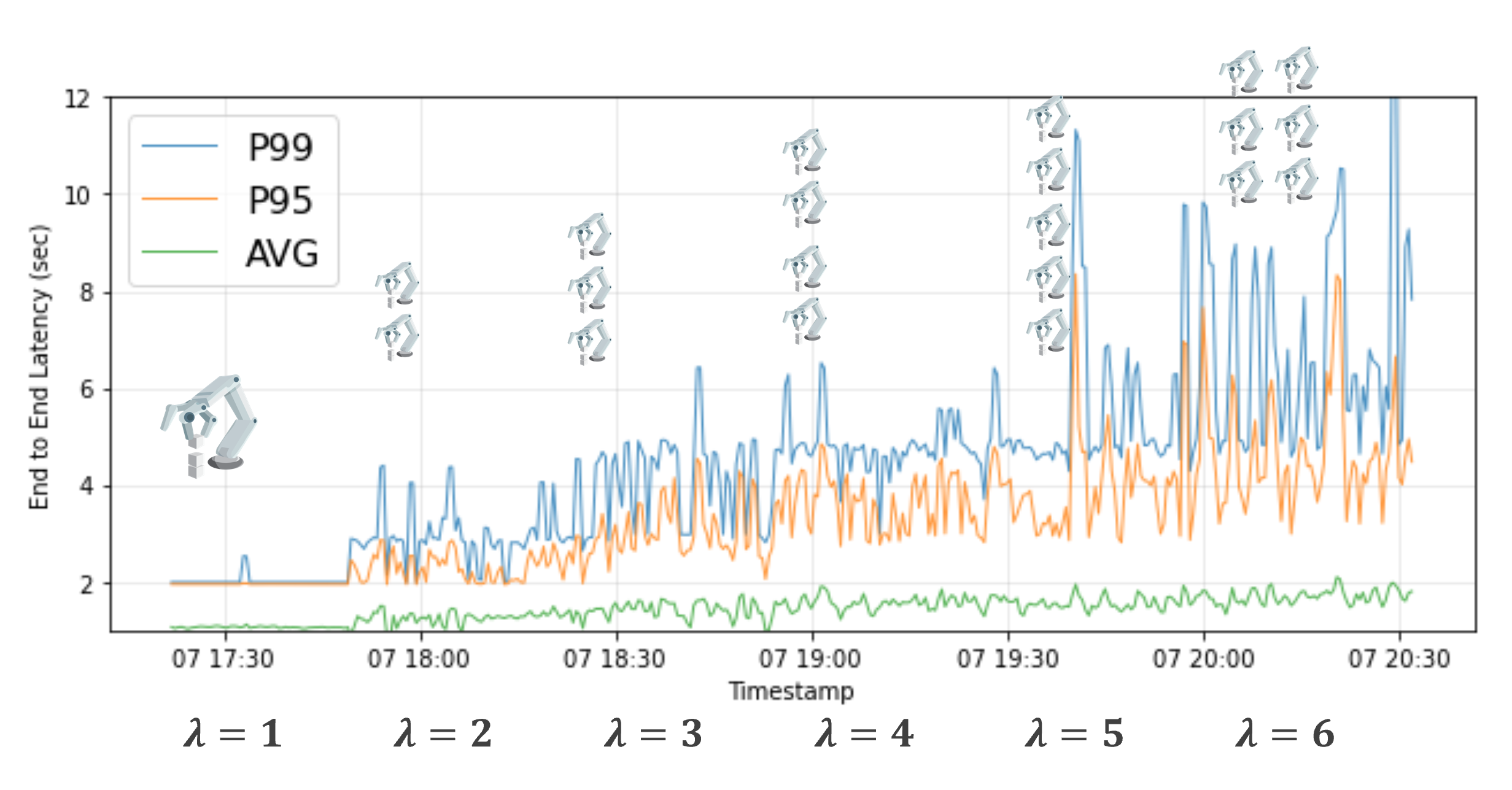}%
}
\caption{Latency comparison of LA-IMR and the baseline latency-based method across varying arrival rates $\lambda \in \{1, 2, 3, 4, 5, 6\}$. LA-IMR significantly reduces tail latencies, particularly the P99 latency, indicating better performance under high load.}
\label{fig:comparision_LA-IMR_baseline}
\end{figure*}

\subsubsection{\textbf{Predictive‑Metric Horizontal Pod Autoscaling (PM‑HPA)}}
We introduce Predictive-Metric Horizontal Pod Autoscaling (PM-HPA), a proactive and latency-aware enhancement to Kubernetes' standard HPA. Rather than relying solely on traditional resource metrics, each microservice computes and exports a single custom metric, desired replicas, based on an internal closed-form queuing model that translates real-time request rates into the optimal number of replicas to mitigate the 99th percentile (P99). Latency measurements are collected via Prometheus, and the computed replica count is exposed to the native HPA, enabling it to scale pods responsively without altering the Kubernetes control plane. PM-HPA responds to traffic surges in milliseconds and maintains graceful shutdowns during scale-in. This results in improved tail-latency performance while maintaining full compatibility with both standard and managed Kubernetes environments.

\subsubsection{\textbf{Experimental Setup and Test Scenario}}
\label{subsubsec:exp_setup}

LA‑IMR runs on a Kubernetes edge cluster hosting a \texttt{YOLOv5m} object‑detection microservice.  
A single CPU replica averages \(L_m^{\text{infer}} \approx 0.8\,\text{s}\); the robot\(\rightarrow\)router\(\rightarrow\)edge\(\rightarrow\)robot round‑trip contributes another \(\approx 1\,\text{s}\).  
Hence the latency SLO is set to
\[
  \tau \;=\; x\,L_m^{\text{infer}} \;=\; 1.8\,\text{s},
\]
with a safety margin \(x = 2.25\) to absorb transient network and queueing delays.

\smallskip
\noindent
Unless stated otherwise, all experiments use the following calibrated parameters:
\begin{itemize}[leftmargin=*]
  \item EWMA smoothing weight: \(\alpha = 0.8\)
  \item Cost–latency trade‑off in~Eq.~\eqref{eq:stageB-obj}: \(\beta = 2.5\)
  \item Utilisation–latency exponent: \(\gamma = 0.90\)
\end{itemize}

To evaluate latency robustness, we steadily increase the arrival rate \(\lambda\)—equivalently, the number of robots issuing requests—while recording the \(\mathrm{P95}\) and \(\mathrm{P99}\) response time.  
Whenever the predicted latency for a replica pool exceeds the SLO, LA‑IMR automatically scales the pool horizontally by incrementing the replica count \(N_{m,i}\) in accordance with Algorithm~\ref{alg:laimr:xscaled}. This closed‑loop reaction keeps the system below the instability boundary and maintains tail‑latency within the configured envelope.

\subsection{\textbf{Latency Evaluation under Workload Fluctuations}}
\label{subsec:eval-fluctuations}

Fig.~\ref{fig:comparision_LA-IMR_baseline} illustrates a comparison between LA‑IMR and a conventional \emph{latency-focused} autoscaling strategy as the incoming request rate~$\lambda$ varies from~1 to~6 requests per second. When the system operates under light load conditions ($\lambda\!\le\!3$), both mechanisms maintain the service-level objective (SLO), exhibiting comparable median response times. However, as the demand rises, the traditional baseline shows noticeable latency variability. Specifically, at $\lambda=6$, the 99th percentile latency (\(\text{P99}\)) reaches 6.8 seconds, whereas LA‑IMR constrains \(\text{P99}\) to no more than 5.4 seconds. This improvement is primarily attributed to its anticipatory scaling of replicas and targeted offloading strategies. As a result, LA‑IMR achieves significantly more consistent tail latency while maintaining similar average performance levels.

\subsection{\textbf{Mitigation of Long‑Tail Latency}}\label{subsec:eval-tail}
\begin{figure}[htbp]
  \centering

  \subfloat[LA-IMR\label{fig:LA_p99_latency_a1_to_a6_box}]{%
    \includegraphics[width=0.25\columnwidth]{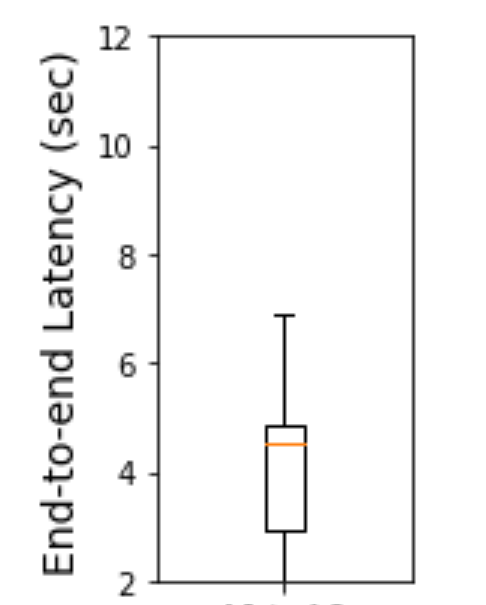}%
  }
  \hspace{0.02\columnwidth}
  \subfloat[Baseline\label{fig:Simple_p99_latency_a1_to_a6_box}]{%
    \includegraphics[width=0.25\columnwidth]{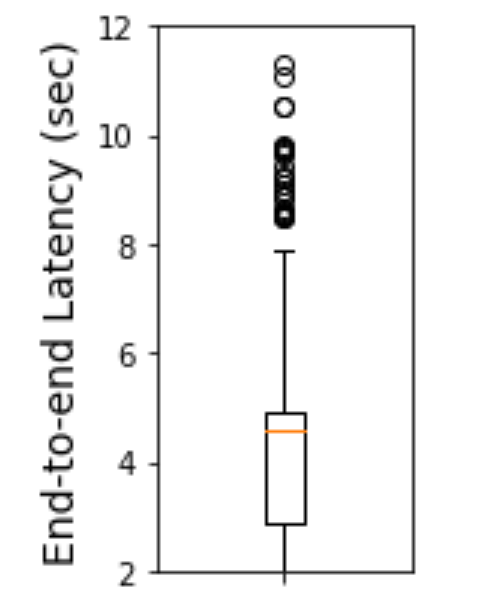}%
  }

  \caption{Box plots of $\mathrm{P99}$ latencies using Prometheus measurements for arrival rates $\lambda=1$--$6~\mathrm{req/sec}$. LA-IMR reduces the interquartile range by 27\% a
  \label{fig:p99_latency_a1_to_a6_box}nd the maximum outlier by 41\%.}
\end{figure}

The Prometheus telemetry in Fig.~\ref{fig:p99_latency_a1_to_a6_box} highlights LA‑IMR’s superior tail‑latency control: its inter‑quartile range is narrower, and extreme outliers are absent.
Algorithm~\ref{alg:laimr:xscaled} predicts queue build‑ups from the closed‑form model and either scales out or off‑loads \emph{before} long queues materialise, thereby suppressing otherwise destructive spikes.

\begin{table}[ht]
  \centering
  \caption{P95 and P99 latencies (mean $\pm$ SD, sec) across arrival rates $\lambda$; lower numbers are \textbf{bold}.}
  \label{table:latency_comparison}
  \begin{tabular}{lcccc}
    \toprule
    & \multicolumn{2}{c}{\textbf{P95}} & \multicolumn{2}{c}{\textbf{P99}}\\
    \cmidrule(lr){2-3}\cmidrule(l){4-5}
    $\lambda$ (req/s) & LA‑IMR & Baseline & LA‑IMR & Baseline\\
    \midrule
    1 & \textbf{1.947$\pm$0.003} & 1.950$\pm$0.004 & \textbf{1.989$\pm$0.001} & 2.012$\pm$0.106\\
    2 & 2.287$\pm$0.568 & \textbf{2.278$\pm$0.288} & \textbf{2.858$\pm$0.826} & 2.933$\pm$0.598\\
    3 & \textbf{2.928$\pm$0.494} & 3.107$\pm$0.566 & \textbf{4.042$\pm$0.856} & 4.201$\pm$0.863\\
    4 & 3.692$\pm$0.703 & \textbf{3.634$\pm$0.575} & \textbf{4.167$\pm$0.902} & 4.782$\pm$0.526\\
    5 & \textbf{3.314$\pm$0.471} & 3.963$\pm$1.091 & \textbf{4.782$\pm$0.639} & 5.632$\pm$1.717\\
    6 & \textbf{4.051$\pm$0.599} & 4.649$\pm$1.125 & \textbf{5.435$\pm$0.827} & 6.855$\pm$2.208\\
    \bottomrule
  \end{tabular}
\end{table}

Table~\ref{table:latency_comparison} shows that LA-IMR consistently achieves lower or equal P95 latency compared to the baseline, with the largest reduction of 14\% at $\lambda=5$~req/s. For P99, the gains grow with load—from 1\% at $\lambda=1$ to 20.7\% at $\lambda=6$—averaging around 9\% overall. At peak load, LA-IMR also cuts the P99 standard deviation by over 60\% (2.21~s~$\rightarrow$~0.83~s), greatly reducing outliers and improving SLO stability.

\subsection{\textbf{Discussion}} \label{subsec:eval-discussion}
\textbf{Experimental setup.} LA‑IMR was evaluated on a shared Kubernetes cluster whose pod start‑up and tear‑down times fluctuate with node availability, image caching, and network contention, introducing real‑world noise that can mask fine‑grained effects.  For tractability we limited the study to two vision workloads—EfficientDet‑Lite0 and YOLOv5m—and tuned the EWMA weight $\alpha$, utilisation floor$\rho_{\text{low}}$, and latency‑budget multiplier $x$ offline for their specific SLOs. Deployments with stricter SLOs or more volatile demand may therefore need adaptive self‑tuning.

\textbf{Limitations and future work.} Load bursts were emulated with a bounded‑Pareto process, whereas real incidents (e.g., holiday shopping) often cause correlated spikes across services.  We also left global off‑loading and cross‑cluster load balancing unoptimised from the cloud‑provider’s perspective—an open problem for future work.

\section{Conclusion}\label{sec:conclusion}
We advance latency‑sensitive edge–cloud inference by coupling a closed‑form latency model—capturing processing, network, and queueing delays—with LA‑IMR, a predictive, SLO‑aware control layer that unites quality‑stratified microservices, event‑driven autoscaling, and selective offloading. Kubernetes experiments show LA‑IMR trims P99 latency by up to 20.7\% and cuts its variance by more than half compared with a reactive autoscaler, owing to prediction‑guided offloading that deflects bursts before queues form and proactive replica provisioning that adds capacity before utilisation nears instability.

We will extend LA‑IMR by incorporating memory‑intensive, variable‑batch workloads to stress‑test its latency model, replacing static control knobs with an online self‑tuner that continuously maximises “SLOs met per dollar,” and combining fast‑ and slow‑window arrival‑rate estimators to catch sudden spikes without destabilising steady traffic.

\section*{Acknowledgment}
\bibliographystyle{IEEEtran}
\bibliography{bibtemplate}

\end{document}